\documentclass[journal,comsoc]{IEEEtran}
\usepackage[T1]{fontenc}

\usepackage{graphicx}     
\usepackage{amsmath}      
\usepackage[sort&compress,numbers,square]{natbib}
\usepackage{epstopdf}
\usepackage[font=small,labelfont=bf]{caption}
\usepackage{xcolor}
\usepackage{bytefield} 
\usepackage{datatool}
\usepackage{glossaries}
\usepackage{booktabs} 
\usepackage{seqsplit}
\usepackage[labelfont=bf]{caption}
\usepackage{blindtext}
\usepackage{float}
\usepackage{amsmath}

\usepackage{graphicx}     
\usepackage{amsmath}      
\usepackage[sort&compress,numbers,square]{natbib}
\usepackage{epstopdf}
\usepackage[font=small,labelfont=bf]{caption}
\usepackage{xcolor}
\usepackage{bytefield}

\usepackage{pgfplots}
\usepackage{pgfplotstable}
\usepackage{caption}
\usepackage{subcaption}
\usepgfplotslibrary{external}  
\usepackage{glossaries} 
\usepackage{graphicx}
\usepackage{tikz}
\usetikzlibrary{calc,positioning,shapes,decorations.pathreplacing}

\pgfplotsset{compat=newest}

   \usepackage[hidelinks=true]{hyperref}

\ifCLASSINFOpdf
\else
\fi
\usepackage{amsmath}
\usepackage[cmintegrals]{newtxmath}
\begin{document}
%
\title{A Novel Approach to Quality of Service Provisioning in Trusted Relay Quantum Key Distribution Networks}
%
%
%

\author{Miralem~Mehic,~\IEEEmembership{Member,~IEEE,}
        Peppino~Fazio,~\IEEEmembership{Member,~IEEE,}
        Stefan~Rass,~\IEEEmembership{Member,~IEEE,}
        Oliver~Maurhart,~\IEEEmembership{Member,~IEEE,}
        Momtchil~Peev,~\IEEEmembership{Member,~IEEE,}
        Andreas~Poppe,~\IEEEmembership{Member,~IEEE,}
        Jan~Rozhon,~\IEEEmembership{Member,~IEEE,}
        Marcin~Niemiec,~\IEEEmembership{Senior~Member,~IEEE,}
        and~Miroslav~Voznak,~\IEEEmembership{Senior~Member,~IEEE,}
        
\thanks{M. Mehic is with Department of Telecommunications, Faculty of Electrical Engineering, University of Sarajevo, Bosnia and Herzegovina and VSB-Technical University of Ostrava, Czech Republic; e-mail: miralem.mehic@ieee.org}
\thanks{P. Fazio is with Department DIMES, University of Calabria, Italy}
\thanks{S. Rass is with the System Security Group, Institute of Applied Informatics,
Alpen-Adria Universit\"at Klagenfurt, A-9020 Klagenfurt, Austria}
\thanks{O. Maurhart is with AIT Austrian Institute of Technology GmbH, Digital Safety \& Security Department, Vienna, Austria}
\thanks{M. Peev and A.Poppe are with Huawei Technologies Duesseldorf GmbH, Munich, Germany}
\thanks{J.Rozhon and M.Voznak are with the VSB-Technical University of Ostrava, Czech Republic}
\thanks{M. Niemiec is with AGH University of Science and Technology, Krakow, Poland}
\thanks{TCOM version based on Michael Shell's bare{\textunderscore}jrnl.tex version 1.3.}}

%
%

\markboth{Journal of \LaTeX\ Class Files,~Vol.~14, No.~8, August~2015}%
{Shell \MakeLowercase{\textit{et al.}}: Bare Demo of IEEEtran.cls for IEEE Communications Society Journals}
%



\maketitle

\begin{abstract}

In recent years, noticeable progress has been made in the development of quantum equipment, reflected through the number of successful demonstrations of Quantum Key Distribution (QKD) technology. Although they showcase the great achievements of QKD, many practical difficulties still need to be resolved. Inspired by the significant similarity between mobile ad-hoc networks and QKD technology, we propose a novel quality of service (QoS) model including new metrics for determining the states of public and quantum channels as well as a comprehensive metric of the QKD link. We also propose a novel routing protocol to achieve high-level scalability and minimize consumption of cryptographic keys. Given the limited mobility of nodes in QKD networks, our routing protocol uses the geographical distance and calculated link states to determine the optimal route. It also benefits from a caching mechanism and detection of returning loops to provide effective forwarding while minimizing key consumption and achieving the desired utilization of network links. Simulation results are presented to demonstrate the validity and accuracy of the proposed solutions.

\end{abstract}


%
\IEEEpeerreviewmaketitle

%
\IEEEpeerreviewmaketitle

\section{Introduction} 
 
\IEEEPARstart{D}{uring} the 30 years since the discovery of the first quantum protocol~\cite{bennett1984quantum}, quantum technology has grown significantly and is rapidly approaching high levels of maturity. The next natural step in the evolution of quantum systems is to study their performance, suitability and convergence with applications used in everyday life. Significant progress in the development of quantum equipment has been reflected through a number of successful demonstrations of QKD networks~\cite{Peev2009,Elliott2007,Xu2009,SasakiM20111,Wang2014,Ciurana2015} but without showing the clear suitability to assess how such networks compete with their classical counterparts under real-life conditions and in real-time traffic. The traffic in these networks was mainly considered with equal importance and it was treated with same priority. While such approach may be acceptable for some applications, it is not suitable for voice, video and collaborative applications. Since not all network traffic is equal, it should not be treated equally. Hence, different applications may have different service requirements with respect to quality of service (QoS). 
This paper addresses the question of using real-time communication in QKD networks by considering QoS mechanisms. The primary goal is to provide an adequate QoS model that includes traffic classification and marking mechanisms, QKD link metrics that can be used to accurately describe the state of the network, and a scalable routing protocol that minimizes the consumption of key material through equitable utilization of network resources.

QKD networks differ from traditional networks in several aspects:

\begin{itemize}


\item 
Although theoretical and pioneering results have been published in the field of quantum repeaters and quantum relays~\cite{Collins2005,Dur1999,Yuan2008}, in practice they remain unachievable with current technology\footnote{The idea behind implementing quantum routers is to use quantum entanglement of photons to communicate over different quantum channels. In short, it means that multiple particles are linked together such that the measurement of one particle's quantum state determines the possible quantum states of the other particles. Even when the particles are separated by a large distance, they still make up a joint quantum system. The fidelity of a quantum state decreases exponentially with the distance of its qubits due to noisy quantum channels~\cite{Salvail2010,VanEnk1998}. In addition, quantum memory is required to implement a quantum repeater according to~\cite{Dur1999,Yuan2008}. Although implementations of quantum memories exist, which can store a qubit for between several milliseconds to one second or even more, this is still too short for practical applications.}\cite{Alleaume2014,Salvail2010}. Therefore, communication is realized in a hop-by-hop~\cite{Peev2009} or key relay manner~\cite{Elliott2007,Sergienko2005}. Both methods rely on the assumption that all nodes along the path between the sender and the receiver must be fully trusted, forming a trusted relay QKD network~\cite{Elliott2002,Marhoefer2007,Peev2009}. 

\item 
Nodes are connected with QKD logical links, referred to below as links, which employ two distinct channels: a quantum channel, which is used for transmission of raw cryptographic keys encoded in certain photon properties, and a public channel, used for verification and processing of the exchanged values. Each quantum channel is always a point-to-point connection between exactly two nodes~\cite{Kollmitzer2010}, while public channels can be implemented as any conventional connection which can include an arbitrary number of intermediate devices~\cite{Dianati2007}.

\item 
The key rate is interconnected with a length of optical fiber such that a longer distance implies a lower key rate due to absorption and scattering of photons~\cite{Alleaume2009,Gisin2002,Salvail2010,Scarani2008b}. Although key rates of 1 Mbps and above have been achieved~\cite{Dixon2010a,korzh2015,SasakiM20111,Wang2012}, such solutions are limited to very short distances. Therefore, both endpoints of the corresponding link implement key buffers (storages) of limited capacity, which are gradually filled at their maximum key rate with the processed cryptographic key, referred to below as key material, and subsequently used for encryption/decryption of data flow~\cite{Kollmitzer2010,Dianati2008a}. Key material denotes the symmetric cryptographic keys that are generated during the QKD process, stored in buffers (storages) and used subsequently for cryptographic operations over user traffic.

\item Without key material, cryptographic operations cannot be performed, and a link can be described as temporarily unavailable~\cite{Elliott2002}. To provide information-theoretically secure (ITS) communication, the key tends to be applied with a One-Time Pad (OTP) cipher and authenticated using an ITS message authentication scheme such as Wegman-Carter when communicating over a public channel~\cite{Abidin2011,Portmann2014a,Wegman1981}. As a result, ITS communication requires more bits of key material than the length of the secured message~\cite{Kollmitzer2010}. The type of encryption algorithm used and the volume of network traffic to be encrypted determines the key storage emptying speed, referred to as the key consumption rate. The key consumption rate denotes the rate of key material being taken from buffers (storages) and used for cryptographic operations. Similarly, the charging key rate (or simply key rate) denotes the rate of new key material generation, that is, the rate of adding new keys to the buffers (storages)~\cite{Elliott2007,Kollmitzer2010,Mehic2015a}.
  
\item 
To meet the requirement to bypass untrusted nodes, in practice, QKD networks are usually deployed as overlay point-to-point networks which exhibit selfish behavior, acting strategically to optimize performance and resulting in dynamic and unpredictable link performance~\cite{Andersen2001,Dianati2008a,Kollmitzer2010,Lee2008,Liu2005}. Overlay networks use existing underlying networks in an attempt to implement a better service; one of their most important features is the independence of the path offered by Internet service providers (ISP).

\end{itemize}

This paper is organized as follows: Section~\ref{section:relatedWork} covers related work, Section~\ref{section:TheSimilaritiesMANETandQKD} points out the significant resemblance between QKD and ad-hoc networks, and Section~\ref{section:qualityOfServiceInQKDNetworks} defines the requirements for providing QoS in QKD networks. In Section~\ref{section:FQKD} we propose a novel QoS model including a GPSRQ routing protocol. The simulation setup is presented in Section~\ref{sec:SimulationSetup}, with Section ~\ref{sec:SimulationResultsAndEvaluation} presenting the evaluation of obtained results. Section~\ref{section:conclusion} concludes the paper.

\section{Related Work}
\label{section:relatedWork}

Given the common assumption that all nodes along a path in a QKD network must be fully trusted~\cite{Elliott2002,Marhoefer2007}, QKD networks were mainly analyzed from two aspects: security and network performance. The idea of passive eavesdropping, in which the adversary may use eavesdropping not to extract information but to redirect the data flow towards a node under their control, has been analyzed in~\cite{Rass2012}. Following a similar idea, stochastic routing has been proposed to avoid deterministic routing which is used in traditional routing protocols~\cite{Le2008a,LeQuoc2007}. Game-theoretic techniques have been used to find an optimal balance between interdependent service quality criteria with distinct performance indicators~\cite{Rass2013}.  
 
QoS in QKD networks deployed previously has been largely neglected, stating that it is somehow achievable without any difficulties. Therefore, the Open Shortest Path First (OSPFv2) routing protocol has been modified to use the amount of key material as the routing metric while ignoring the performance of the public channel~\cite{Elliott2005,Dianati2008a}. In~\cite{Tanizawa2016}, the author proposes using unencrypted and non-authenticated communication for the dissemination of OSPFv2 routing packets; this is simple prey for an eavesdropper who is assumed to have unlimited resources at their disposal, especially when passive eavesdropping is taken into account~\cite{Rass2012}. In~\cite{SUN2014,XianzhuCheng2011}, unmodified OSPFv2 was combined with reserving key material resources. Considering the interdependence of the public and quantum channels~\cite{Mehic2017a}, reservation of key material resources does not solve the problem of QoS since the routing path may be inadequate for quantum channels.
  
\section{Similarities Between QKD and MANET Technologies}
\label{section:TheSimilaritiesMANETandQKD} 

The specific QKD issues and constraints described above pose significant challenges in QKD network design. However, by analyzing the characteristics of QKD networks, we note similarities with Mobile Ad Hoc Networks (MANET)~\cite{Fazio2015,Fazio2016,Rango2014}. First, we specify the main characteristics of QKD technology from a simple point of view:
\begin{itemize}
\item QKD links, described above, due to features of quantum channels, are always implemented in point-to-point behavior, and they can be roughly characterized by two features: limited distance and key rate inversely proportional to the distance~\cite{Kollmitzer2010}. Additionally, QKD links may become unavailable when there is no enough key material or when the public channel is congested~\cite{Mehic2017a}. Such behavior is similar to Wi-Fi links which are limited in length and where the communication speed depends on the user's distance from the transmitter.
\item One of the main features of QKD networks is the absence of a quantum repeater or quantum router in practice, therefore communication is usually performed on a hop-by-hop basis~\cite{Alleaume2014,Kollmitzer2010}. 
\end{itemize}

In MANET, communication takes place on a hop-by-hop basis and mobile nodes are typically powered by batteries, placing special attention on energy-aware solutions. The nodes connect themselves in a decentralized, self-organizing manner with no authority in charge of managing and controlling the network~\cite{Rango2012,Fazio2016}. The battery power in MANET nodes can be easily linked to the amount of key material in QKD key storages. Given the node without a power supply (empty batteries) is not an active member of the network, the same analogy is valid for QKD networks where the node without available key material cannot be used for data transmission. The range limitations of wireless links can be mapped to the limitations in the length of QKD links, while the absence of a dedicated network infrastructure (such as a router) is common to MANET and QKD. Here we recognize the significant similarity between these two technologies, allowing us to propose a new approach to addressing QKD network issues.
 
\section{QoS in QKD Networks}
\label{section:qualityOfServiceInQKDNetworks}
 

The specific QKD constraints described above lead to the conclusion that this type of network provides weak support to QoS. 

\subsection{QoS Models}

\subsubsection{Integrated Service and QKD Networks}

The basic concept of the Integrated Service (IntServ) model is per-flow resource reservation using the Signaling Resource ReSerVation Protocol (RSVP) before data transmission~\cite{Zhang1993}. In our view, the IntServ model is not suitable for QKD networks because of the following:
\begin{itemize}
 
\item 
\textit{An inability to guarantee the reservation:} when QoS is provided in an IP network, an IP router has complete control over its packet buffers and the output link bandwidth, and can directly schedule these resources. In contrast, in an overlay network, the node cannot directly access the available resources in the overlay path. It can only rely on measurement techniques where high accuracy cannot be guaranteed and it cannot directly control or reserve resources in the underlying network. The only thing that a node in a QKD network is able to do is to guarantee resources of a quantum channel by reserving key material in key storages. However, considering the interdependence of public and quantum channels in a QKD link~\cite{Mehic2017a}, such reservation does not constitute any gain.

\item \textit{Signaling:} RSVP is an out-of-band signalization protocol which means that signaling packets contend for network resources with data packets and consume a substantial amount of scarce key material and network resources.

\end{itemize}


\subsubsection{Differentiated Service and QKD Networks}

Differentiated Service (DiffServ) uses Differentiated Services Code Point (DSCP) bits in the IP header and a base set of packet forwarding rules known as Per-Hop-Behavior. DiffServ is known as an edge-provisioning model and it does not provide any QoS guarantees \textit{per se}~\cite{Park2005}. The application of DiffServ in its original form is limited in a QKD network due to the following:

\begin{itemize}

\item \textit{Edge router selection}: existing QKD technology limits the deployment of a QKD network to the metropolitan scale\footnote{In 2014, a QKD system connecting the cities of Hefei-Chaohu-Wuhu (HCW) in China with a total of nine QKD nodes was reported in~\cite{Wang2014}. Thirteen QKD devices within nine nodes in total were employed to support the two QKD networks and the intercity QKD link.}~\cite{Ciurana2014,Wang2014,SasakiM20111,Chen2010,Peev2009,Elliott2007}. In such a network, it is necessary to clearly define the edge routers which play a key role in the processing of traffic.

\item \textit{Lack of Service Level Agreement}: each network node needs to comply with the rules for the classification and processing of traffic of different priorities. However, since the Service Level Agreement (SLA) concept is not defined in a QKD network, it is questionable how nodes of potentially different domains can negotiate traffic rules.

\end{itemize}

\subsection{QoS Signalization in QKD Networks}
\label{sec:QoSSignaling}

In general, the QKD protocol which establishes a new key material consists of six successive stages: a physical exchange of quantum states between a pair of devices, extraction of the raw key (sifting), error rate estimation, reconciliation, privacy amplification, and authentication~\cite{Bennett1992a,Mehic2015}. Only the first stage is performed over a quantum channel; all other stages are performed over a public channel, resulting in communication referred to as QKD post-processing. We propose an extension of authenticated packets with signaling data which provides an elegant way of tackling the problem of distribution of signaling information without introducing additional traffic overheads~\cite{Mehic2017b}. 
  
\subsection{QoS Routing in QKD Networks}
\label{sec:QoSRouting}

In our view, a routing protocol well-suited for operation in dynamic QKD networks should meet the following main design objectives:

\begin{itemize}

\item Reducing the consumption of scarce key material by choosing the shortest path considering both channels of the QKD link. The routing algorithm needs to find a balance between the requirements, since a path that meets the requirements of the public channel may not be suitable for the quantum channel and vice versa~\cite{Mehic2017a,Kollmitzer2010}.

\item Given that the main objective of QKD is to provide ITS communication, routing packets need to be encrypted and authenticated~\cite{Kollmitzer2010,Maurhart2009}. This means that the number of routing packets needs to be minimized to preserve scarce key material.
 
\item To prevent denial of service, it is necessary to minimize knowledge about the utilized routing path by reducing the broadcast of routing packets~\cite{Rass2012}.

\item The routing protocol should be scalable to different network sizes.

\item Due to a low key charging rate and overlay networking mode, link interruptions are common in QKD networks. Hence, the routing protocol should be robust enough to find an adequate replacement path.

\end{itemize}
 
In general, routing solutions can be divided into three broad categories: source, hierarchical and distributed routing. The performance of source routing algorithms relies on the availability of precise link state information, while the dynamic nature of QKD networks makes the available link state information inherently imprecise. Given that the constant maintenance of link state information is mostly done by periodic flooding, this solution is inadequate for QKD networks. Although several examples exist of hierarchical network organization~\cite{Xu2009,SasakiM20111,Wang2014}, in our opinion such organization is not suitable for QKD networks since nodes of upper hierarchical levels represent a potentially easy target for attack to disassemble the network. In distributed routing, the computation of the path is shared among network nodes on a periodic basis (proactive) or only when a routing path is requested (reactive). Proactive routing protocols mainly use the static update period time for keeping routes up-to-date, which contradicts the dynamic nature of QKD networks. Therefore, in overlay networks reactive routing performs better in terms of efficiency and stability than proactive routing~\cite{Zhu2006,Andersen2003}. 
 
\section{FQKD: A Flexible Quality of Service Model for QKD Networks}
\label{section:FQKD}

To overcome problems of providing QoS in a dynamic environment, we present a flexible QoS model for QKD networks (FQKD). Our model avoids a centralized resource management scheme or reservation of resources mechanism. Instead, we turn to a distributed approach to control traffic loads by providing soft-QoS constraints without flow or session state information maintained in support of end-to-end communication. FQKD defines three roles for nodes in a QKD network: ingress, interior and egress. Each node can take any of these roles depending on the position in network flow. A source node which sends data is referred to as an ingress node. Interior nodes are nodes that forward data to the final destination node which is referred to as an egress node.

\begin{figure}[H]
  \begin{center}
\includegraphics[width=0.5\textwidth]{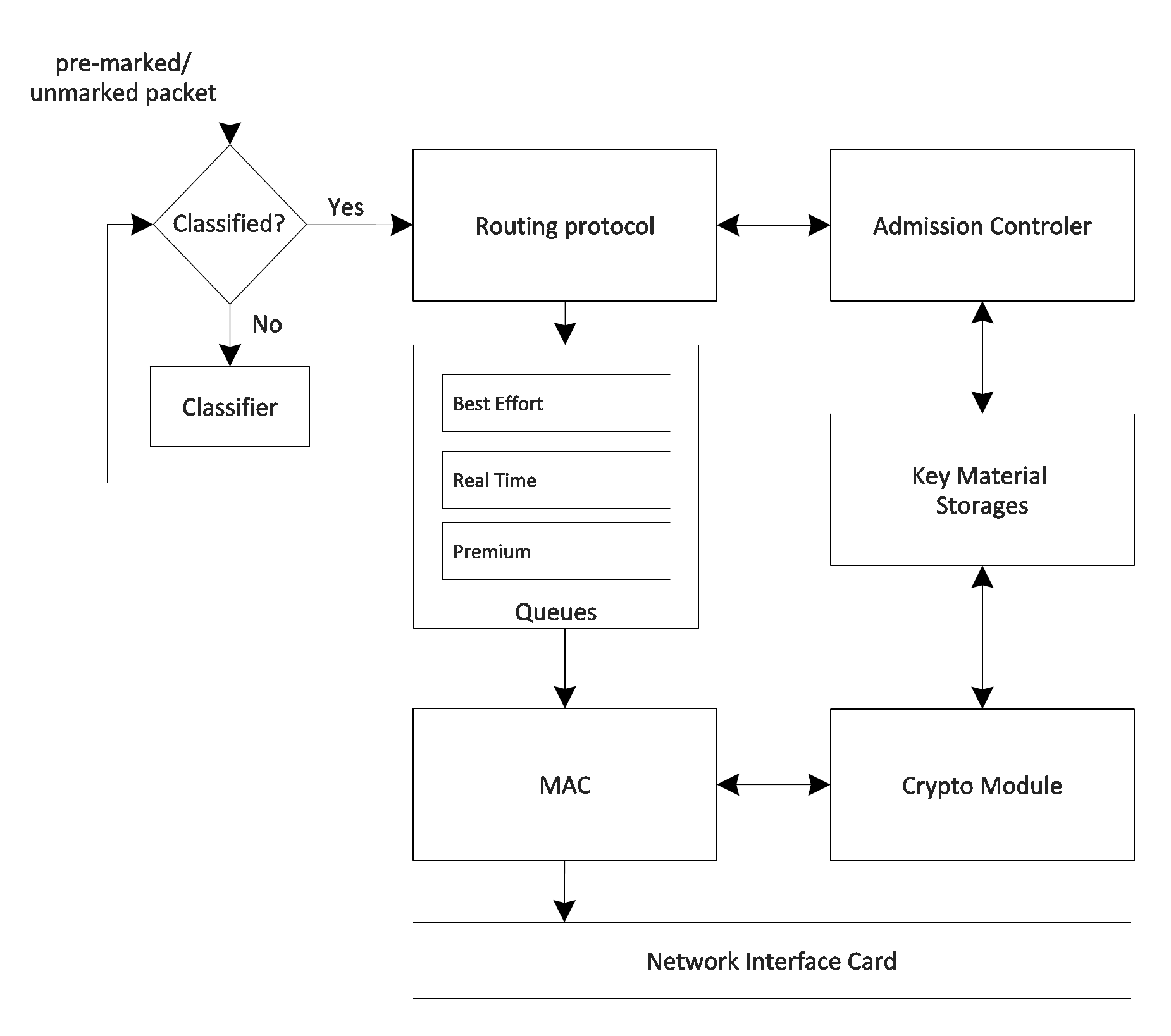}
  \end{center}
  \caption{\small FQKD model. First, the incoming application package is classified and forwarded to the routing protocol which contacts the admission control to verify there are sufficient resources for processing the data. If true, the packet is stored in the appropriate waiting queue and encrypted before sending to the network. Arrows denote communication between model elements and the flow of the processed packet.}
  \label{Fig:FQKD_Model}
\end{figure}
 
\subsection{Provisioning and Conditioning} 
\label{sec:ProvisioningAndConditioning} 
   
As shown in Fig~\ref{Fig:FQKD_Model}, the FQKD model consists of a sender-based classifier, waiting queues, local node-based admission controller, crypto module and dynamic regulation of admitted sessions at the MAC layer. The classifier is input at the ingress node to distinguish between traffic classes by marking the DSCP field in the IP packet header. FQKD distinguishes between three traffic classes with corresponding DSCP values: best-effort, real-time and premium class. For each class, separate waiting queues are defined and processed by priority. The packets are forwarded by interior nodes in per-hop behavior associated with the assigned DSCP value.

Considering that nodes in QKD networks constantly generate new keys at their maximum rate\footnote{QKD nodes aim to generate as much key material as possible. Therefore, nodes constantly generate traffic at the highest rate allowed by the network. More information about the impact of public channel states on the intensity of the key rate can be found in~\cite{Mehic2017a}.}\cite{Dianati2008a}, before setting the route, the routing protocol contacts the admission controller to filter those links to its neighbors that have sufficient resources to serve the classified network packet. The routing protocol calculates the path and the packet is forwarded to the MAC layer for further processing. Otherwise, if no available link is found, the packet waits in the queue for reprocessing. In FQKD, additional waiting queues are installed between the L3 and L4 ISO/OSI layer to avoid conflicts in decision making which could lead to inaccurate routing. Suppose the queues are implemented on the data link layer (L2) only, and suppose that they are half filled with packets. Since the routing protocol used the routing metric that at the time $t_{1}$ of calculating the route had a different value from the time $t_{2}$ when the packet came online in the queue to be served, it follows that significant changes to the state of links in the time interval $\Delta t=t_{2}-t_{1}$ could occur, which can lead to inaccurate and incorrect routing. Instead, by exploiting higher level waiting queues, the packet for which the route is calculated will be directly forwarded to lower layers and immediately sent to the network. This implies the usage of one set of waiting queues (set of three waiting queues for best-effort, real-time and premium traffic classes) for all network interfaces. Using queuing at the L2 layer is not excluded, but additional attention is given to queues at a higher level due to the dynamic nature of the network.

\begin{figure}[!hbt]
  \begin{center}
\includegraphics[width=0.5\textwidth]{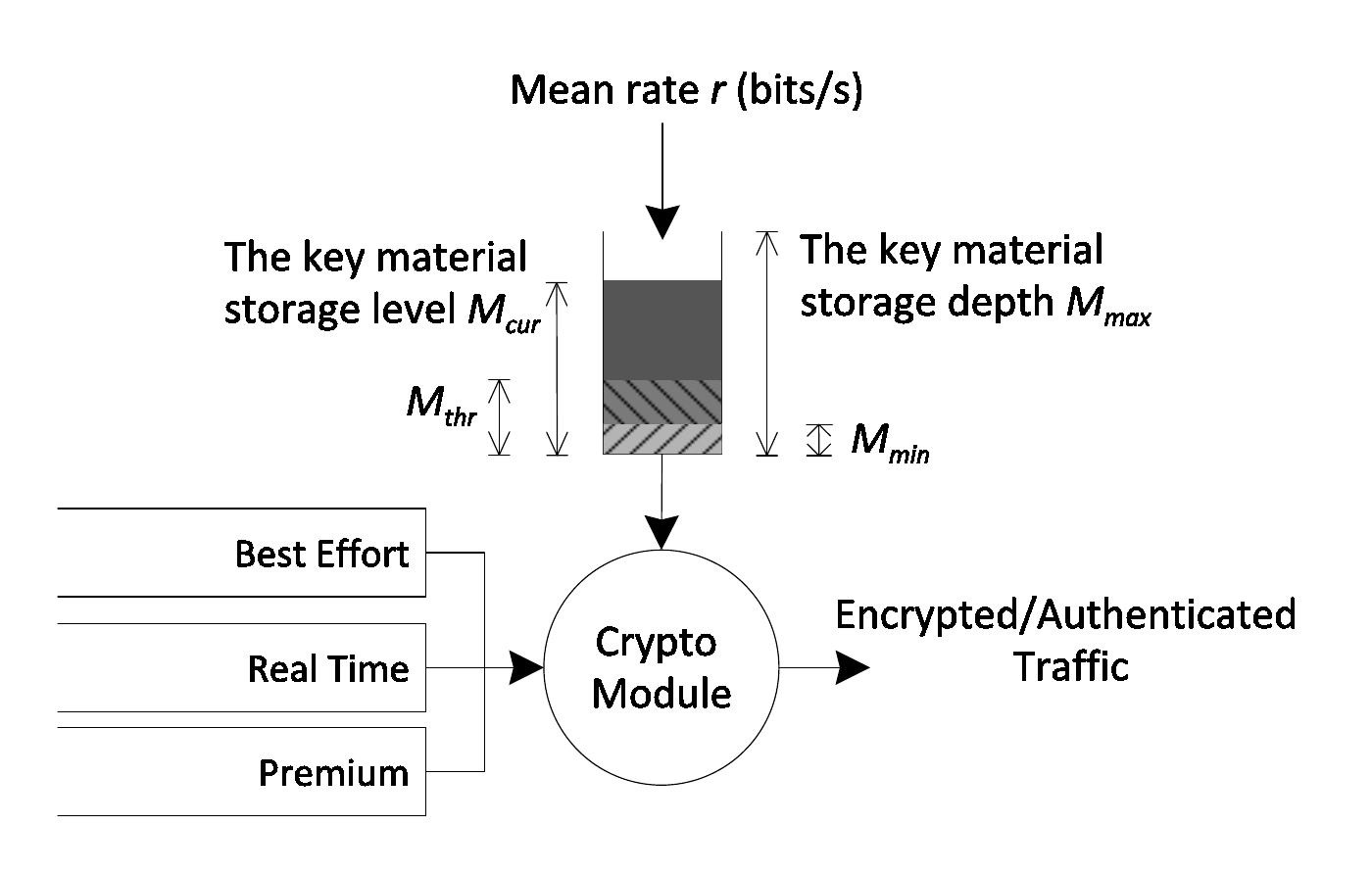}
  \end{center}
  \caption{\small Traffic processing in the FQKD model}
  \label{Fig:TokenBucketmodel}
\end{figure}

Assuming the key rate is constant in time when the quantum channel has a fixed length~\cite{Dianati2008a,Kollmitzer2010}, it is evident that the key storage can be identified with the Token Bucket traffic shaping mechanism as shown in Fig.~\ref{Fig:TokenBucketmodel}. This simplifies the view of the admission controller, which behaves as the traffic conditioner. The volume of traffic over the QKD link is limited by the amount of key material in key storage which is used for encryption or authentication of data over that link. The key material storage of link \textit{k} between nodes \textit{a} and \textit{b} can be represented using following parameters:

\begin{itemize}

\item The time measurement moment \textit{t},

\item The average key generation rate $r_{k}$, measured in bits per seconds and used to indicate the charging rate of the storage,

\item The key material storage depth $M_{max,k}$, used to indicate the capacity of the storage,

\item The current value $M_{cur,k}(t)$, representing the amount of key material in the storage at the time of measurement \textit{t}, where it holds that $\ M_{cur,k}(t)\leq{M_{max,k}}$. 

\item The threshold value $M_{thr,a,b}(t)$ or simply $M_{thr,k}(t)$ described in more detail below,

\item QKD is also known as Quantum Key Growing~\cite{cederlof2008,dodson2009updating} or Quantum Key Expansion~\cite{Pascazio2010}, since it needs a small amount of key material pre-shared between parties (denoted with $M_{min,k}$) to establish a larger amount of the secret key material. This pre-shared key material is used for QKD post-processing and authentication~\cite{Wegman1981,Pascazio2010}.

\end{itemize}

The amount of key material over a time interval $T$ depends on the key generation rate and the available amount of key material in the storage, and can be calculated using Equation~\eqref{eq:tokenBucketEquationAmountOfKeyMaterial}. The average operational rate, that is the rate at which packets can be served over an interval $T$, can be calculated using Equation~\eqref{eq:tokenBucketEquationRate}. As $T\rightarrow\infty$, the operational rate $A_{k}(T)$ approaches the charging key rate $r_{k}$.

\begin{equation} \label{eq:tokenBucketEquationAmountOfKeyMaterial}\begin{aligned}
D_{k}(T)\leq{r_{k} \cdot T+M_{cur,k}(T)-M_{min,k}(T)}
\end{aligned} \end{equation}%
\begin{equation} \label{eq:tokenBucketEquationRate} 
A_{k}(T)=\frac{D_{k}(T)}{T}=r_{k}+\frac{M_{cur,k}(T)-M_{min,k}(T)}{T}
\end{equation}
 
However, the overall amount of traffic data that can be transmitted over link $k$ in a time interval $T$ can be calculated by dividing the value obtained from Equation~\eqref{eq:tokenBucketEquationAmountOfKeyMaterial} and the ratio $L_{k}$ which is the quotient of key length used for encryption with authentication and the length of the data message. Note that for ITS communication, which usually involves the OTP cipher and Wegman-Carter authentication, more bits of key material than the length of the data message are required ($L_{k}>1$)~\cite{Kollmitzer2010}.

An incoming packet is served from the queue only if there is enough key material in the storage. Otherwise, the packet remains in the queue waiting for storage to be charged. The length of the queue is limited and traffic-shaping algorithms are in charge of packet management operations. To avoid blocking work due to the lack of the key material used to generate new key material, special attention is placed on the categorization of the traffic. If storage stops charging, the purpose of the link loses functionality. Therefore, the traffic generated by the post-processing application has the highest priority and is sorted in the premium queue. Only traffic from post-processing applications can use the key material when $\ M_{cur,k}(t)\leq{M_{min,k}}$ while the traffic from the other two queues is served only when $\ M_{cur,k}(t)>{M_{min,k}}$.  
 
\setlength{\belowcaptionskip}{0pt}
\begin{figure}
  \begin{center}
\includegraphics[width=0.42\textwidth]{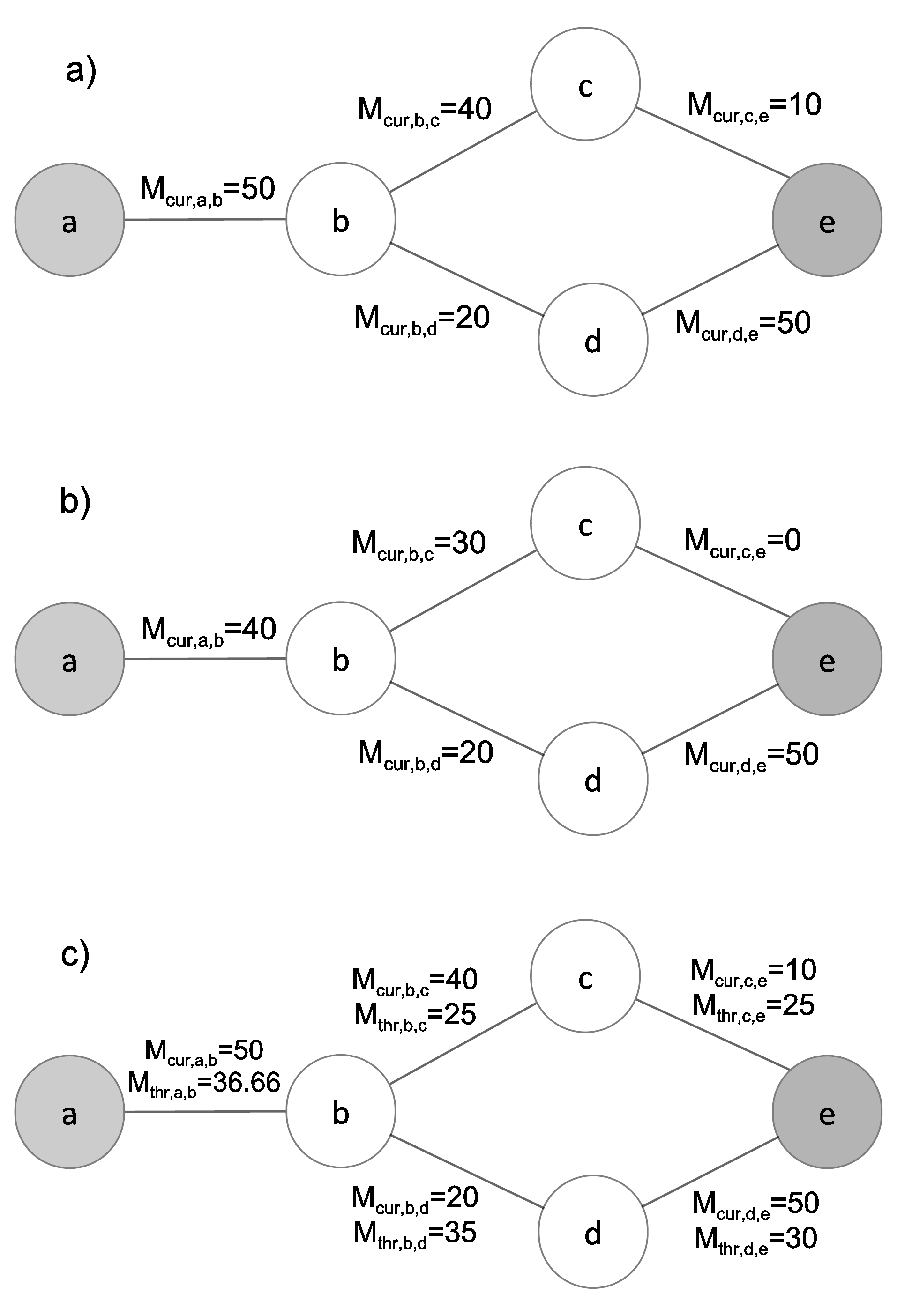}
  \end{center}
  \caption{\small Simple topology showing the calculation of $\ M_{thr}$: a) The traffic is routed along the route \textit{a-b-c-e}; b) The traffic is routed along the path \textit{a-b-c} regardless of key material depletion of link \textit{c-e}; c) 
Calculated $\ M_{thr}$ values based on $\ M_{cur}$ values of all links connected to a given node }
  \label{Fig:Mthr_example}
\end{figure}

The threshold value$\ M_{thr}$ is proposed to increase the stability of QKD links, where$\ M_{thr,k}(t)\leq{M_{max,k}}$. The parameter is explained by considering the simple topology shown in Fig.~\ref{Fig:Mthr_example} where node \textit{a} needs to communicate with remote node \textit{e}. Suppose the routing protocol uses only information about the state of its links with its neighbors. Then, assuming that all network links have the same performance of public channels, we consider only the state of key storages which are marked next to links as shown in Fig.~\ref{Fig:Mthr_example}-a. Upon receipt of the packet from node \textit{a}, the routing protocol on node \textit{b} selects path \textit{b-c} since the link \textit{b-d} has a lower performance. However, if node \textit{b} does not consider the state of links that are more than one hop away, the traffic may stuck on the link between nodes \textit{b} and \textit{c} as shown in Fig.~\ref{Fig:Mthr_example}-b. To avoid such behavior, we propose using the$\ M_{thr}$ value. Each node \textit{i} calculates value$\ L_{i}$ summarizing the$\ M_{cur}$ values of links to its neighbors and dividing it by the number of its neighbors$\ N_{i}$ using Equation~\eqref{eq:Mthr1}. Then, each node exchanges calculated value$\ L_{i}$  with its neighbors. The minimum value denoted by Equation~\eqref{eq:Mthr2} is accepted as a reference threshold value of the link.
 
\begin{equation} \label{eq:Mthr1}
L_{i} =  \frac{\sum_{k=0}^{N_{i}}M_{cur,i,k}}{N_{i}}
\end{equation}%
\begin{equation} \label{eq:Mthr2}
M_{thr,i,j} =  min\{L_{i},L_{j}\}
\end{equation}
 
As shown in Fig.~\ref{Fig:Mthr_example}-c, node \textit{b} calculates$\ L_{b}=36.66$, while node \textit{c} calculates$\ L_{c}=25$. The threshold value of the link \textit{b-c} is set to$\ M_{thr,b,c}=25$ and it is included in link metric calculation as described in Section~\ref{sec:QKDLinkMetric}. The higher the value of $\ M_{thr}$, the better the state of links that are more than one hop away.
\subsection{QKD Link Metric} 
\label{sec:QKDLinkMetric}

Popular metrics from conventional networks which describe the state of the communication link cannot be adequately used in QKD networks since they only describe the public channel. Therefore, we propose new metrics that clearly define the state of the QKD link taking into account its most important features.%

~\subsubsection{The Quantum Channel Status Metric}
 
At the moment of serving the packet, the remaining key material in the key storage is the main factor contributing to the link's availability; this is because without key material, cryptographic operations cannot be performed and secure communication over the link is not possible. We use Equation~\eqref{qc2} to express the state of the quantum channel between nodes \textit{s} and \textit{i}, where $Q_{frac,s,i}$ is the ratio of the squared amount of key material at the time of measurement (${M_{cur,s,i}}^2$) multiplied by the threshold value ($M_{thr,s,i}$) and the cubed capacity of the key storage (${M_{max,s,i}}^3$) as defined by Equation~\eqref{qc1}. $Q_{frac,s,i}$ is in the range [0,1] and it highlights the current amount of key material on closest links in relation to the amount of key material of links that are further away, since more distant links are unreachable when links to neighbors are unavailable.

\begin{equation} \label{qc1}
Q_{frac,s,i} =  \frac{{M_{cur,s,i}}^2 \cdot M_{thr,s,i} }{{M_{max,s,i}}^3}
\end{equation}

\begin{equation} \label{qc2}
Q_{m,s,i} =  1-\frac{Q_{frac,s,i}}{e^{(1-Q_{frac,s,i})} }
\end{equation}

$Q_{m,s,i}$ is the utility function associated with the key material level of the link. $Q_{m,s,i}$ uses an exponential formula to address the fact that the lower the amount of key material in the storage, the more critical the situation, and the less time is left for the routing protocol to react.

\begin{figure}[!ht]
  \begin{center}
   \resizebox{0.9\linewidth}{!}{
		\includegraphics[width=0.42\textwidth]{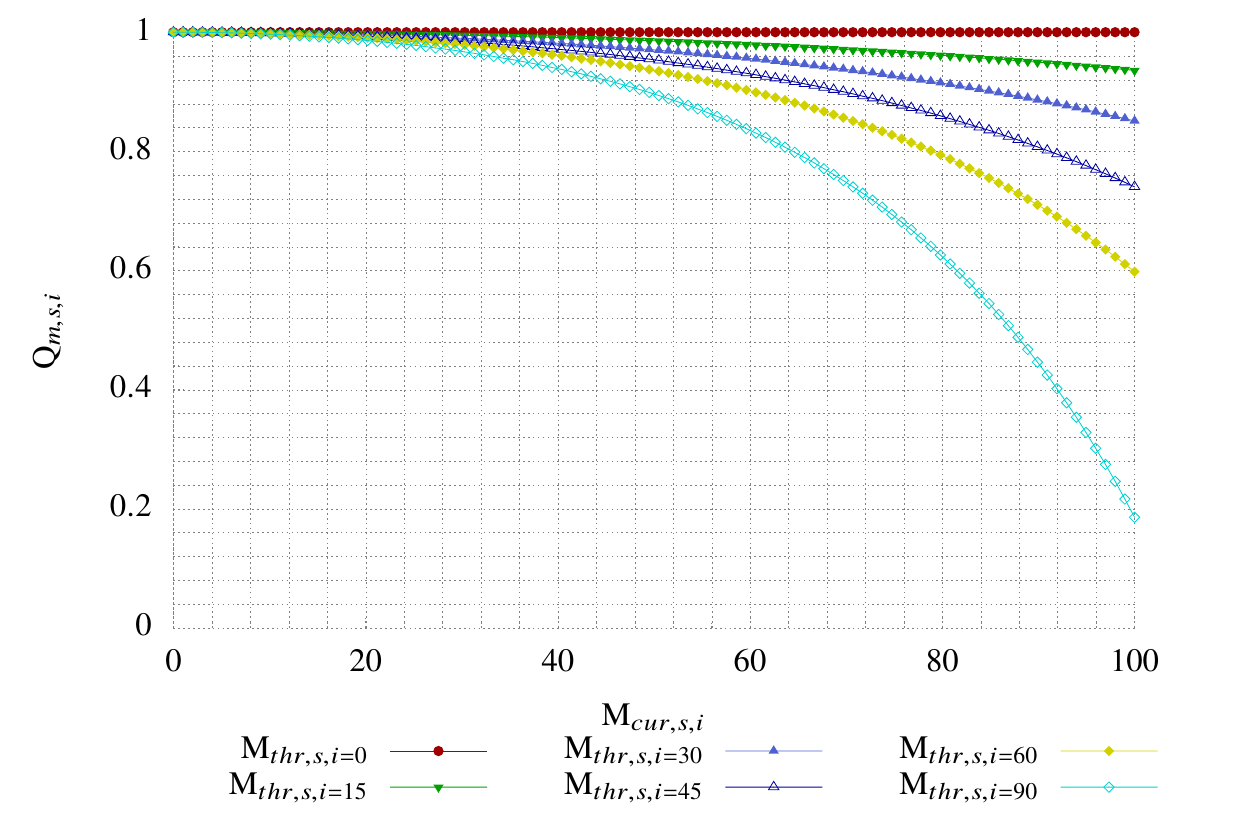}
	}
  \end{center}
  \caption{\small $Q_{m}$ of link between nodes $s$ and $i$ for different values of $M_{thr}$; $M_{max,s,i} = 100$}
  \label{fig:QfracVsMthreshold}
\end{figure}

The value of $Q_{m,s,i}$ is normalized as a grade ranging from 0 to 1 as shown in Fig.~\ref{fig:QfracVsMthreshold}, where a lower value means a better quantum channel state. In the example shown in Fig.~\ref{Fig:Mthr_example}-c, the routing protocol should favor the link \textit{b-d} since $Q_{m,b,d}<Q_{m,b,c}$.

~\subsubsection{The Public Channel Status Metric} 
\label{sec:publicChannelRoutingMetric}

Instead of using popular approaches from conventional or overlay networks~\cite{Andersen2001, Mehic2016}, we use meta-data of keys such as the time duration of the key establishment process to effectively assess the state of the public channel. We define $P_{m,s,i}$ with Equation~\eqref{eq:publicChannel2} to evaluate the state of the public channel between nodes \textit{s} and \textit{i} where $T_{last,s,i}$ is the length of time spent on the establishment of the key material at the time of measurement and $T_{maximal,s,i}$ is the maximum time that can be tolerated for the establishment of the key. $T_{maximal,s,i}$ is calculated as the double value of the average duration of key material establishment process in the long run, denoted as $T_{average}$ in Equation~\eqref{eq:publicChannel1}. 
 
\begin{equation} \label{eq:publicChannel2}
P_{m,s,i} =  \frac{T_{last,s,i}+\Delta t}{T_{maximal,s,i}}
\end{equation}
\begin{equation} \label{eq:publicChannel1}
T_{maximal,s,i} = 2 \cdot T_{average}
\end{equation}

$\Delta t$ is used to describe the freshness of the information and is defined as the difference between the current time of the measurement and the time when the $T_{last,s,i}$ is recorded. Note that $T_{average}$ is not equal for all links of the network, since it depends on the load of the network, types of quantum and network devices, QKD post-processing application and the state of the public channel\footnote{In practice, the QKD link employs three channels. The QKD quantum-channel and the QKD distillation-channel are usually referred to as the quantum channel and public channel, respectively. The third channel is known as the (time-stable) synchronization channel and cannot be separated from the quantum channel physically to two fibers, because it delivers time references. From the perspective of higher network layers, the synchronization channel is part of the key generation (growing) process. If the impact of this channel on the entire process were to be considered, the signals in the synchronization channel need to be sufficiently large so that even co-propagating classical channels or spoiled signals are irrelevant and key generation is disturbed. Such major non-synchronization would be reflected in the overall duration of the key generation process, which is considered using Equation~\eqref{eq:publicChannel2}.
}. The value of$\ P_{m,s,i}$ mainly falls in the range [0,1] where the lower value means a better public channel state. Values greater than 1 indicate that the link has a problem with establishing new key material and such links should not be considered.
 
~\subsubsection{The Overall QKD Link Status Metric}
\label{sec:OverallQKDLinkStatusMetric}

Key material depletion is not the same for all links since it depends on the type of encryption algorithm used and the volume of network traffic to be encrypted. For example, a QKD link between nodes \textit{s} and \textit{i} having a low value of $Q_{m,s,i}$ may be suitable for a network flow encrypted with less secure algorithms which require less key material than OTP (such as the AES cipher), but it may not be suitable for encryption using OTP. Thus, factor $\alpha$ which reflects the balance between the requirements is introduced in Equation~\eqref{eq:final_routing_metric} to compensate for this effect using the utility functions of the quantum and public channel in the [0,1] range where a lower value means a better overall link state.%

\begin{equation} 
\label{eq:final_routing_metric}
R_{m,s,i} = \alpha \cdot Q_{m,s,i} + (1-\alpha) \cdot P_{m,s,i}
\end{equation}%
Parameter $\alpha$ takes the value in the [0,1] range; if the OTP cipher is used, we suggest the value of $\alpha=0.5$. This means that both channels of the QKD link are considered equally. If the AES cipher is used, $\alpha$ can be set to a lower value to put a greater emphasis on the public channel due to its lower requirements for key material, depending how frequently it is refreshed and the length of the AES key used.

\section{Greedy Perimeter Stateless Routing Protocol for QKD network} 
\label{sec:GPSRQ}

Driven by the similarities between the MANET and QKD networks discussed in Section~\ref{section:TheSimilaritiesMANETandQKD}, we present the Greedy Perimeter Stateless Routing Protocol for QKD networks (GPSRQ). The main motivation for designing GPSRQ is to minimize the number of routing packets and to achieve high-level scalability by using distributed geography reactive routing. %

We assume that all nodes know the geographical locations of all other network nodes they wish to communicate with. Therefore, there is no periodic flooding of the node location details and we assume a location registration and lookup service that maps the node address to a location. This paper does not deal with implementation details of such a service, but we assume it can be implemented using internal or other communication channels~\cite{Li2000}. As indicated in Section~\ref{sec:QoSSignaling}, authenticated packets in QKD post-processing can be used to effectively exchange information about the geographical position of nodes. Although several experiments have been conducted regarding mobile QKD networking~\cite{Wijesekera2011,Sheikh2006,Schmitt-Manderbach2007,Vallone2014,Yin2017}, due to the high sensitivity of quantum equipment to various environmental factors\footnote{The key rate may vary due to humidity, temperature, stability of devices, global radiation, pressure, dust, sunshine duration or other factors~\cite{Dusek2006,Kollmitzer2010}.} and relatively low key generation rates in free-space QKD links~\cite{Liao2017,Pugh2016,korzh2015,shimizu2014,Pascazio2010}, in this paper we assume that QKD networks are composed of static nodes representing secure access points\footnote{Low mobility within a geographic region or slow node movement are supported as long as the service for distributing geographic locations of nodes can accurately and safely distribute locations to all nodes in the network.}. We follow the idea "\textit{the greater the distance separating two nodes, the slower they appear to be moving with respect to each other}" outlined in~\cite{Basagni1998} to implement caching in GPSRQ, which is discussed further below.

GPSRQ sets up a network without hierarchical organization, which means that all nodes in the network are of equal importance. Nodes do not exchange routing tables, which significantly minimizes the consumption of scarce key material and reduces the probability of passive eavesdropping~\cite{Rass2012}. Route selection, that is the decision about the next hop, is made in per-hop behavior such that the packet is moved closer to the destination based on the states of links in the local environment and on the geographical distance from the node. An eavesdropper is not able to intercept routing packets and find out the exact route to the destination, since it is not known at which node's network interface the packet will be forwarded until the last moment. GPSRQ uses two packet forwarding algorithms: greedy forwarding and recovery mode forwarding.

\begin{figure}[!ht]
  \begin{center}
  \includegraphics[width=0.4\textwidth]{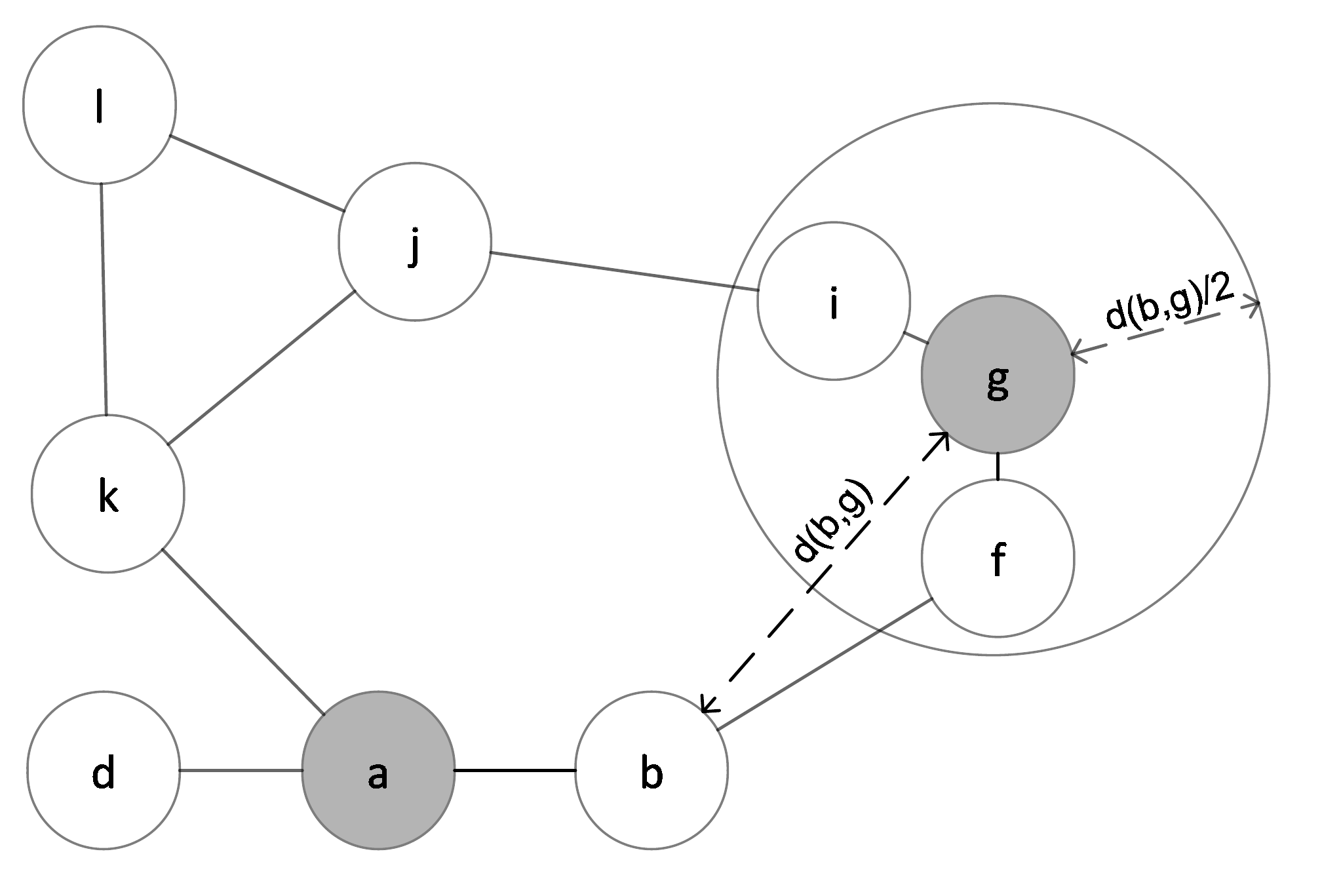}
  \end{center}
  \caption{\small Ingress node \textit{a} which is surrounded by three adjacent nodes \textit{b}, \textit{d}, and \textit{k}, aims to communicate with the egress node \textit{g}. In the absence of a path over node \textit{b}, node \textit{a} writes in the internal cache that it is not possible to route packets along the path \textit{a-b} towards the region marked with a circle of radius \textit{d(b,g)/2} with the center in node \textit{g}. Any further request for routing towards any node which is placed in the defined circle region will be ignored over the route via node \textit{b}, and an alternative route needs to be found. }
  \label{fig:fqkd_graph_with_cricle}
\end{figure}
  
\subsection{Greedy Forwarding} 
\label{sec:GreedyForwarding}
 
By definition, greedy forwarding entails forwarding to the neighbor geographically closest to the destination. An example of greedy forwarding is shown in Fig.~\ref{fig:fqkd_graph_with_cricle}, where ingress node \textit{a} which is surrounded by three adjacent nodes \textit{b}, \textit{d}, and \textit{k}, needs to communicate with egress node \textit{g}. Ingress node \textit{a} forwards the packet to \textit{b}, as the Euclidean distance between \textit{b} and \textit{g} is smaller than the distance between \textit{g} and any of \textit{a}'s other neighbors. Such greedy forwarding is repeated on interior nodes and it stops when the packet reaches its destination. GPSRQ aims to maximize network utilization by using different paths for different traffic classes. It uses Equation~\eqref{eq:FQKD_QoS_Equation} to calculate the path to forward the packet:
\begin{equation} \label{eq:FQKD_QoS_Equation}
F_{a,g,v} = \big(1-\beta\big) \cdot R_{a,v} + \beta \cdot G_{v,d}
\end{equation}
where $R_{a,v}$ denotes the state of link between source node \textit{a} and neighboring node \textit{v} using Equation~\eqref{eq:final_routing_metric} and $G_{v,g}$ represents the Euclidean distance between neighboring node \textit{v} and destination node \textit{g}, for each node \textit{v} which belongs to the set $N_{a}$ of all neighbors of source node \textit{a}, $\forall v \in N_{a}$. All routes towards the destination are sorted in descending order using Equation~\eqref{eq:FQKD_QoS_Equation} and the route with the lowest value is used. GPSRQ uses parameter $\beta$ with a value in the [0,1] range to manage network utilization by choosing between forwarding along the geographically shortest route or the route that has the most available resources. We discuss simulation results for different values of $\beta$ in Section~\ref{section:effectOfBetaParameter}. 
 
Greedy forwarding relies on the knowledge of the geographical location and state of links to neighbors, which results in a high level of network scalability. However, in some cases the route to the destination requires forwarding the packet over a neighboring node which is geographically further from the destination than the node which forwards the packet. In cases when a local maximum occurs, an alternative recovery-mode forwarding is used. To increase scalability and exclude routes which do not lead to the destination, we propose a robust caching mechanism to preserve key material consumption. Consider the example shown in Fig.~\ref{fig:fqkd_graph_with_cricle} where we further suppose that link \textit{b-f} is unavailable due to a lack of key material. When node \textit{b} realizes there is only one interface available (the interface which was used to receive the packet from node \textit{a}), it marks a "loop" field in the GPSRQ packet header and returns the packet back to node \textit{a}. Then, node \textit{a} calculates the Euclidean distance \textit{d(b,g)} between node \textit{b} and destination node \textit{g} and writes in its internal cache memory that it is not possible to route towards the region which is marked with a circle of radius \textit{d(b,g)/2} and the center in node \textit{g} along the path \textit{a-b}. Upon receiving any further requests for routing towards the node placed in the defined circle region, node \textit{a} will ignore the route over node \textit{b} and look for an alternative route. The validity of cached record is set to time interval defined using Equation~\eqref{eq:t_cache}:\begin{equation} 
\label{eq:t_cache}
T_{cache}= T_{maximal,b,g}/2
\end{equation}
where $T_{maximal,b,g}$ is defined by Equation~\eqref{eq:publicChannel1}. The overall goal is to reduce the dynamics of the network topology changes using the scalable cache mechanism. In Section~\ref{section:effectOfPublicChannelMemory}, we discuss the impact of different values on cache validity. After the cached record expires, the node is allowed to try establishing the connection once again. In the event when GPSRQ detects there is no neighbor closer to the destination, it enters recovery mode.

\subsection{Recovery Mode Forwarding} 
\label{sec:RecoveryModeForwarding}

Recovery mode involves using the well-known right-hand rule which states that the next edge from node \textit{k} upon arriving from node \textit{a} is edge \textit{(k,j)} which is sequentially counterclockwise to edge \textit{(a,k)}~\cite{Karp2000}. However, for first forwarding, the packet is forwarded along edge \textit{(a,k)} which is counterclockwise to node \textit{a} from line $\overline{ag}$. The packet stays in recovery mode until it reaches a node which is closer to the destination than the node where forwarding in recovery mode started. To avoid routing loops, the GPSRQ header contains information about the IP address of the node at which the packet entered recovery mode and the outgoing interface which was used for first forwarding. If the loop is detected by analyzing the packet header, the packet is returned to the previous node for rerouting and adding a new entry to the node's internal cache memory. If the public channel of link \textit{j-i} is unavailable, there is no available path to destination \textit{g}. Source node \textit{a} forwards the packet to node \textit{k} which forwards the packet to node \textit{j} in greedy forwarding mode. Since link \textit{j-i} is unavailable, node \textit{j} is not able to find any neighbor closer to the destination so it enters recovery mode, sets the value of the field "inRec" to 1, writes its IP address to the header field "recPosition", writes the interface number which leads to node \textit{l} in the "recIF" header field and forwards the packet to node \textit{l} since it is on the first edge counterclockwise about \textit{j} from the line $\overline{jg}$ as required by the right-hand rule. Upon receipt of the packet, node \textit{l} inspects the header and remains in recovery mode since node \textit{j} in which the packet entered recovery mode is closer to the destination \textit{g}. Then, the packet is forwarded to node \textit{k} which forwards the packet back to node \textit{j}. When node \textit{j} detects its IP address from the header field "recPosition", it adds a record to the internal cache memory stating that it is not possible to reach destination node \textit{g} via node \textit{l} by adding a record consisting of the triple: IP address of hop \textit{l}, the radius of circle region \textit{d(l,g)/2} and the value of the circle center which is set to the location of node \textit{g}. Given that there is no other interface available, node \textit{j} sets GPSRQ header field "loop" to 1 and returns the packet to node \textit{k} which
adds a record to its cache memory stating that it is not possible to reach destination node \textit{g} via node \textit{j}. Then, node \textit{k} sets the field "loop" to 2 and tries again with greedy forwarding where node \textit{j} is excluded as the next hop. The packet is forwarded to node \textit{l} which forwards the packet to node \textit{j}. Since node \textit{j} does not have any other interface available, it will set the value of header field "loop" to 1, and return the packet to node \textit{l} which will add to its cache memory a record stating that it is not possible to reach destination node \textit{g} over node \textit{j}. The packet is returned to node \textit{k} and then it is returned to node \textit{a}. This procedure is repeated until a feasible path to the destination is found. Otherwise, if no available path is found, the packet is gradually returned to the source node which is allowed to discard the packet while keeping updated records in the cache memory of other nodes.

\subsection{GPSRQ Protocol Implementation}
\label{sec:gpsrq_protocol_implementation}

\begin{table}[!t]
\caption{GPSRQ packet header fields used in recovery mode forwarding}
\label{table:gpsrq_header}
\centering
\begin{tabular}{|c|c|}
\hline
Field & Description \\
\hline
inRec & Mode: Greedy or Recovery \\
\hline
recPosition & IP address of the node where recovery mode started \\
\hline
loop  & Returning Loop indicator   \\
\hline
recIF & Interface used for forwarding in recovery mode     \\
\hline
\end{tabular}
\end{table}

\begin{figure}[!ht]
  \begin{center}
  \includegraphics[width=0.4\textwidth]{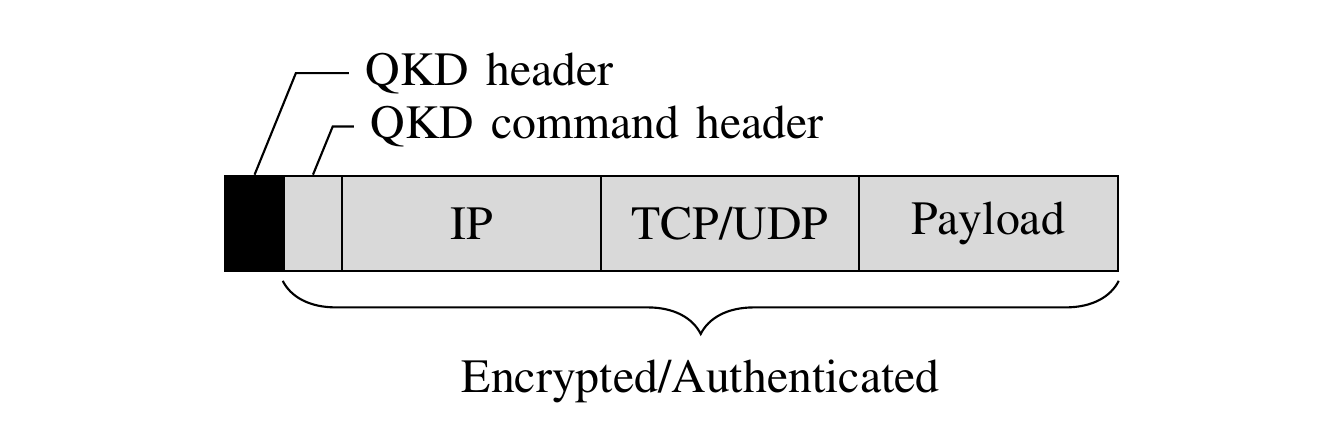}
  \end{center}
\caption{Packet encapsulation with the QKD header and QKD command header}%
\label{fig:qkdCommandHeaderpacketEncapsulation}%
\end{figure}

To make the routing operation easier, encryption and authentication are performed on the data link ISO/OSI layer~\cite{Kollmitzer2010,Mehic2016a}. However, in previously deployed QKD networks the QKD header was transported unencrypted~\cite{Maurhart2009a}. A denial-of-service attack is possible if an eavesdropper is able to block high importance packets such as routing packets or post-processing operation packets which would disable the link~\cite{Mehic2016b} or divert traffic to the node under an attacker's control~\cite{Rass2012}. Therefore, we propose using the QKD Command header which is encrypted together with the packet payload (Fig.~\ref{fig:qkdCommandHeaderpacketEncapsulation}). Instead of adding the GPSRQ header to the packet, which would increase the amount of key used for encryption, values from the GPSRQ header which are listed in Table~\ref{table:gpsrq_header} are rearranged to the QKD header and QKD command header prior to transmission. Upon receiving the packet, the values are moved from the QKD header and QKD command header to the GPSRQ header.

In typical multimedia communication protocols, excessively delayed packets are not used for the reconstruction of transmitted data at the receiver~\cite{Sun2013a}. We propose the extension of the QKD header by including \textit{Timestamp} and \textit{MaxDelay} fields to carry the packet's timestamp and maximum tolerated delay, respectively. The values of these fields are written at the ingress node, while the GPSRQ at the interior node checks the values before forwarding the packet, including the previously mentioned field "loop". If the value of the "loop" field is equal to 2, the packet was in a loop which was later avoided and marked in the internal cache in the previous nodes, so no action is taken against the packet. If the value of the "loop" field is equal to 0, it means that the packet was not previously in the loop. Then, GPSRQ checks whether the delay of the packet, which is calculated as the difference between the current timestamp and the Timestamp value, is greater than the MaxDelay. If this is true, GPSRQ sets the "loop" field to 1 and returns the packet to the previous hop which will result in adding a new entry to the internal cache in previous nodes as mentioned earlier. The function of the Timestamp and MaxDelay fields is similar to the Time-to-Live (TTL) field in the IP header of conventional networks, but with the aim of minimizing the consumption of scarce key material in QKD network.

\section{Simulation Setup}
\label{sec:SimulationSetup}

To ensure simulations are independent of the topology or characteristics of any specific network, random graphs were constructed. Random network topologies were generated using the Waxman model~\cite{Waxman88} which is recommended for small and medium size networks which include locality aspects such as QKD networks~\cite{Ciurana2014,Wang2014,Peev2009}. Additionally, the Waxman model corresponds to the requirement for the implementation of QKD networks without hierarchical multi-plane organization~\cite{Pussep2010} since it spreads nodes randomly on a grid and adds links randomly, such that the probability $P_{e}$ of interconnecting two nodes in a single plane is parameterized by the Euclidean distance separating them, as defined using Equation~\eqref{eq:waxman}:

\begin{equation} \label{eq:waxman}  
P_{e}(u,v) = \Theta \cdot exp^{-\frac{d(u,v)}{\Omega \cdot \Lambda }} 
\end{equation}

where $d(u,v)$ is the Euclidean distance between nodes $u$ and $v$, $\Lambda$ denotes the maximum possible distance between two nodes where $0<\Omega, \Theta \leq 1$. The parameter choices are constrained to assure $P_{e}(u,v)\in[0,1]$. Each simulation was repeated four times for random network topologies, with the same number of nodes, and with random values of the initial amount of key material in key storages. We evaluated GPSRQ ($\beta=0.6$; $T_{average}=5$) against OSPFv2 which was used in previously deployed QKD networks~\cite{Tanizawa2016,Dianati2008a,Kollmitzer2010,SUN2014,XianzhuCheng2011}, and against DSDV which was used in our previous work~\cite{Mehic2016b}. The simulation was performed using the QKD Network Simulation Model (QKDNetSim)~\cite{Mehic2017} to deploy GPSRQ and DSDV while NS-3-DCE v.1.9 was used to deploy the OSPFv2 routing protocol~\cite{Riley2010}. We used the BRITE topology generator to generate random topologies according to the Waxman model, since it is supported under NS-3 and the source code is freely available~\cite{Medina2001}. NS-3-DCE and QKDNetSim were set to share the same seed file to generate random values which enabled us to use the same random topologies with the same configuration values in QKDNetSim and NS-3- DCE. Simulations included static networks with 10, 20, 30, 40, 50, 60 and 70 nodes which were randomly placed in a rectangular region and connected to QKD links with the following settings: minimal amount of key material 1 Mbyte; maximal amount 100 Mbyte; initial amount randomly generated in the [0.5, 25] MByte range; maximal bandwidth of the link set to 10 Mbps; charging key rate set to 100 kbps with the charging key period of seven seconds. The Waxman router model was used with the following parameter values: $m$ = 2; $\Lambda$ = 100; $\Omega$ =  0.4; $\Theta$ = 0.4; The first node was set as the source of traffic, while the last randomly placed node was set as the destination. The source node generated UDP traffic with a 1 Mbps rate and fixed packet size of 512 bytes which were encrypted using OTP and authenticated using VMAC with a 32 bit authentication tag~\cite{Abidin2011,Portmann2014a,Wegman1981}. The duration of the simulation was 150 seconds, while the capacity of waiting queues per device on the L2 and L4 layers (used for GPSRQ only) was set to 1000 packets. The parameters not given here are the default parameters of QKDNetSim and NS-3-DCE.

\subsection{Simulation Results and Evaluation}
\label{sec:SimulationResultsAndEvaluation}

Although GPSRQ can achieve valuable results on non-planar graphs, one of the drawbacks of GPSRQ is the large consumption of key material in non-planar networks since geographical routing cannot quickly determine the shortest path towards the destination which leads to unnecessary forwarding. It is, therefore, advisable to convert non-planar graphs into planar graphs in which geographic routing protocols can be effectively used. Instead of using the heuristic to exclude the intersecting edges~\cite{Karp2000a}, we modified the BRITE random topology generator to generate random planar Gabriel Graphs\footnote{The BRITE random generator with Gabriel Graph module is available at \textit{https://bitbucket.org/mickeyze/brite-planar-graph}} (GG) which are defined as follows: "An edge "\textit{An edge $(u,v)$ exists between vertices $u$ and $v$ if no other vertex $w$ is present within the circle whose diameter is the Euclidean distance d(u,v)}"~\cite{Chandy1982b,Gabriel1969}.

\subsubsection{Routing Protocol Overhead}
\label{sec:routingProtocolOverhead} 
 
\begin{table*}[t]
\centering
\footnotesize
\caption{The number and sizes of generated routing packets}
\label{table:routingPacketSizes}
\begin{tabular}{|l|l|l|l|l|l|l|}
\hline
 & \multicolumn{2}{l|}{GPSRQ} & \multicolumn{2}{l|}{DSDV} & \multicolumn{2}{l|}{OSPFv2} \\ \hline
Nodes & \begin{tabular}[c]{@{}l@{}}Number of\\ Packets\end{tabular} & \begin{tabular}[c]{@{}l@{}}Sum of Packet  \\ Sizes (MByte)\end{tabular} & \begin{tabular}[c]{@{}l@{}}Number of\\ Packets\end{tabular} & \begin{tabular}[c]{@{}l@{}}Sum of Packet \\ Sizes (Mbyte)\end{tabular} & \begin{tabular}[c]{@{}l@{}}Number of\\ Packets\end{tabular} & \begin{tabular}[c]{@{}l@{}}Sum of Packet \\ Sizes (MByte)\end{tabular} \\ \hline
10 & 1740 & 0.097 & 4832 & 0.417 & 1594 & 0.149 \\ \hline
20 & 4256 & 0.238 & 41 k & 3.208 & 5522 & 0.628 \\ \hline
30 & 6638 & 0.372 & 115 k & 8.582 & 11 k & 1.422 \\ \hline
40 & 7896 & 0.442 & 142 k & 11 & 13 k & 1.829 \\ \hline
50 & 9996 & 0.56 & 283 k & 20 & 18 k & 2.618 \\ \hline
60 & 12198 & 0.684 & 294 k & 23 & 27 k & 3.866 \\ \hline
70 & 14804 & 0.831 & 604 k & 44 & 36 k & 5.468 \\ \hline
\end{tabular}
\end{table*}

\begin{figure}[htbp]
  \begin{center}
  \includegraphics[width=0.4\textwidth]{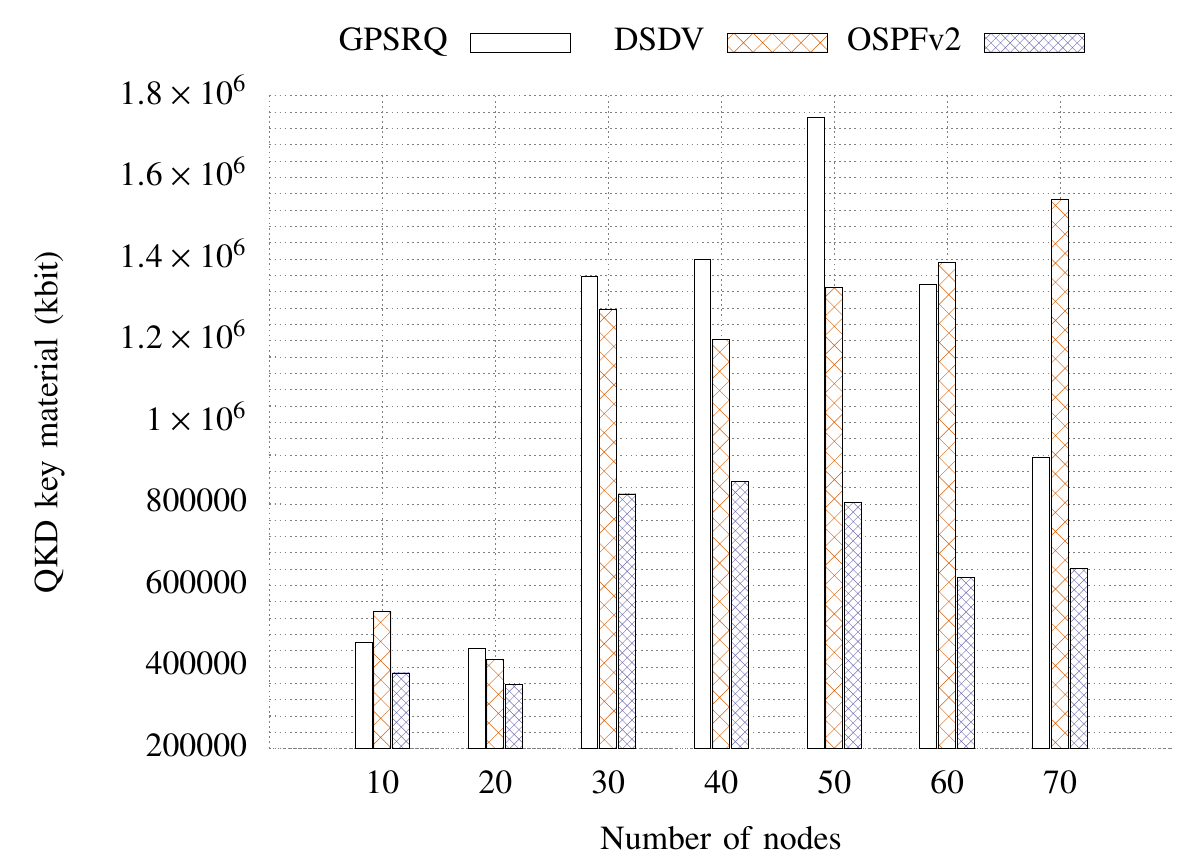}
  \end{center}
  \caption{\small Average Key material consumption}
  \label{fig:qkd_key_consumption_graph} 
\end{figure}

Table~\ref{table:routingPacketSizes} shows the routing protocol overhead, measured by the number and summed sizes of routing packets sent network-wide during the simulation for GPSRQ, DSDV and OSPFv2. OSPFv2 is a widely deployed link-state routing protocol which uses the periodic Link State Announcement (LSA) flooding mechanism to update link-state databases describing the network topology~\cite{Moy1991}. By default, OSPFv2 floods LSA update information every 30 minutes and it exchanges Hello packets to establish and maintain a neighbor relationship every 10 seconds. If a node does not receive a Hello message from a neighbor within a fixed dead interval of time which is set to a default of 40 seconds for point-to-point networks, OSPFv2 modifies its topology database entries to indicate that the neighbor is unavailable. DSDV is a proactive routing protocol which periodically broadcast its routing table to its neighbors (every 15 seconds by default). In addition, DSDV uses triggered updates when the network topology suddenly changes. If periodic and triggered updates occur in a short period of time, the values may be merged and only the periodic update will be performed~\cite{Perkins1994,Mehic2016}.
In contrast, GPSRQ relies on the knowledge of the geographical position and state of links to neighbors, which provides a high level of network scalability. As such, it periodically exchanges only $M_{thr}$ packets defined in Section~\ref{sec:ProvisioningAndConditioning} every time a new key material is stored in key storages; the value was set to seven seconds in our simulations. It is important to note that DSDV and OSPFv2 exchange their routing packets using UDP, while GPSRQ exchanges $M_{thr}$ values using TCP, so Table~\ref{table:routingPacketSizes} includes all packets including TCP SYN, TCP ACK and TCP FIN. We performed several simulations where $M_{thr}$ was set to 5, 15, 20 and 30 seconds besides the exchange of $M_{thr}$ every time new key material is added to key storage. The data showed the same values, letting us conclude that a single exchange of a $M_{thr}$ value in a QKD post-processing period is adequate for the smooth operation of GPSRQ. In our simulations, values from GPSRQ packets were moved into the QKD command header as described in Section~\ref{sec:gpsrq_protocol_implementation}, while DSDV and OSPFv2 packets were sent using the standard QKD header~\cite{Kollmitzer2010}. Table~\ref{table:routingPacketSizes} shows that GPSRQ consumes the lowest amount of key material for cryptographic operations in routing packets, while Fig.~\ref{fig:qkd_key_consumption_graph} shows that GPSRQ consumes the greatest amount of the available key material, but for securing data traffic as seen from the PDR value (Section~\ref{sec:packetDeliveryRatio}).

 \begin{figure}[htbp]
  \begin{center}
  \includegraphics[width=0.4\textwidth]{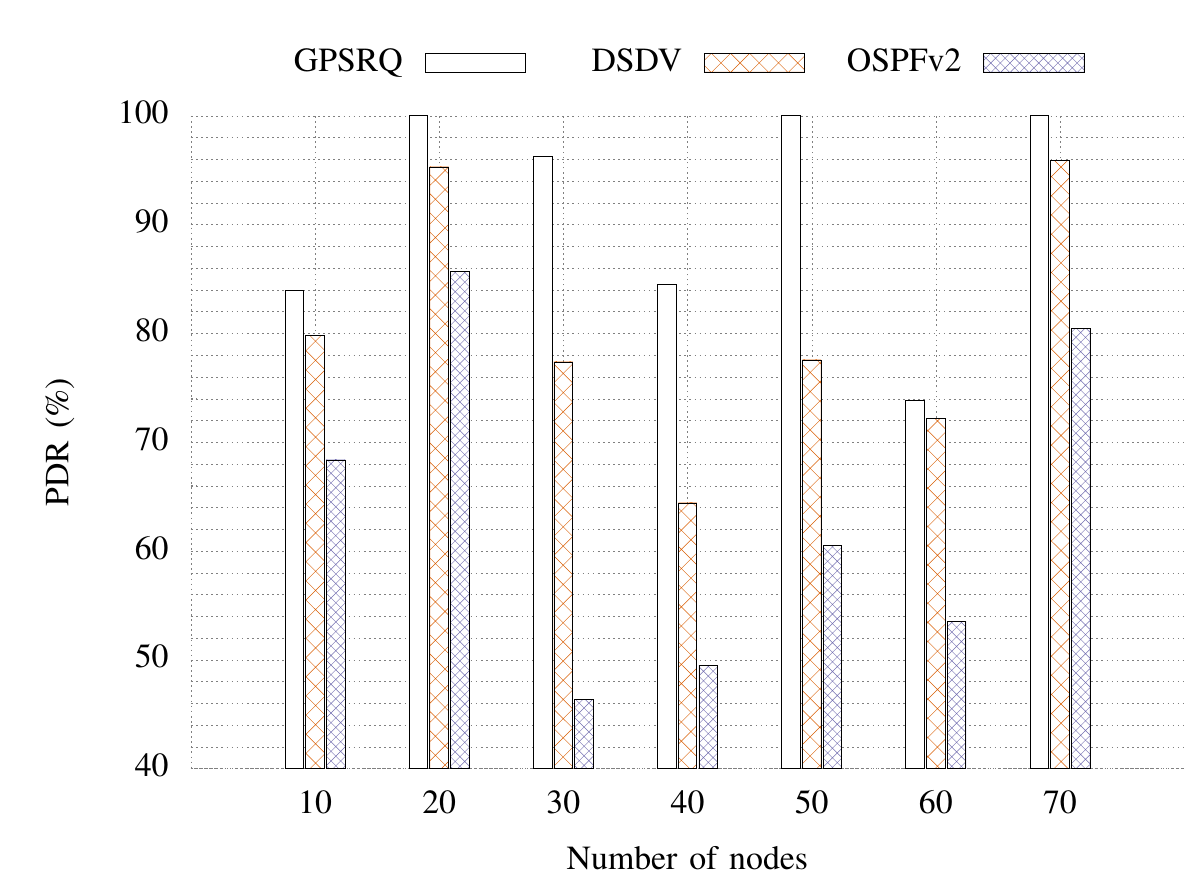}
  \end{center}
  \caption{\small Average Packet Delivery Ratio}
  \label{fig:non_planar_pdr} 
\end{figure}

\subsubsection{Packet Delivery Ratio}
\label{sec:packetDeliveryRatio}

The Packet Delivery Ratio (PDR), calculated as the ratio of received and sent application packets, is used to assess the effectiveness of the routing protocol within the specified simulation environment. Fig.~\ref{fig:non_planar_pdr} shows that GPSRQ is able to successfully find an available route to a destination when compared to OSPFv2 and DSDV. Due to the large value of the dead interval, OSPFv2 is not able to react quickly to the changes in network topology which results in a reduced PDR value. DSDV exchanges routing packets more often; this returns a higher PDR value, although it is still significantly lower than GPSRQ. It is important to note that in some simulation scenarios PDR could not reach 100\% because it depends on the network conditions as defined by random values, such as the initial amount of key material in key storages. However, using the same seed files ensures identical network conditions for all tested protocols.

 \begin{figure}[htbp]
  \begin{center}
  \includegraphics[width=0.4\textwidth]{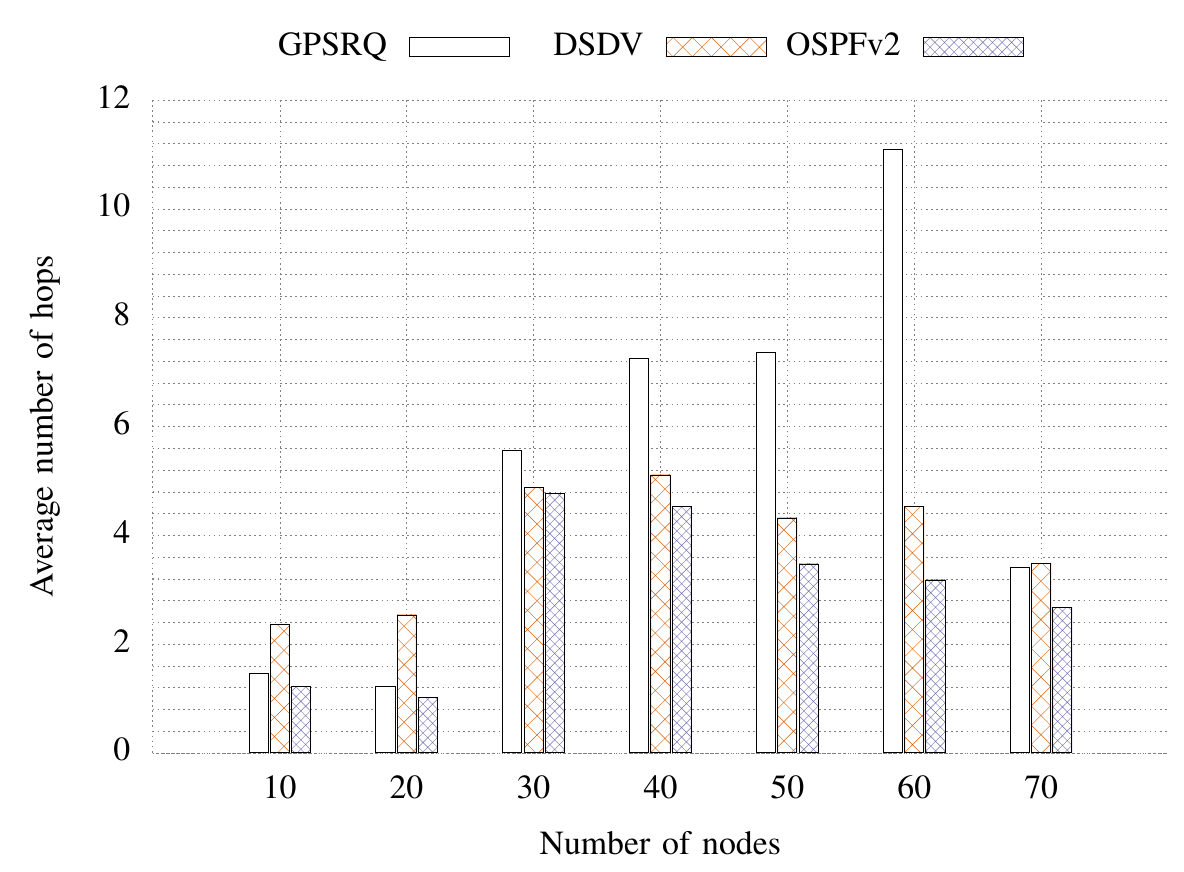}
  \end{center}
  \caption{\small Average number of hops}
  \label{fig:planar_pdr} 
\end{figure}

 \begin{figure}[htbp]
  \begin{center}
  \includegraphics[width=0.4\textwidth]{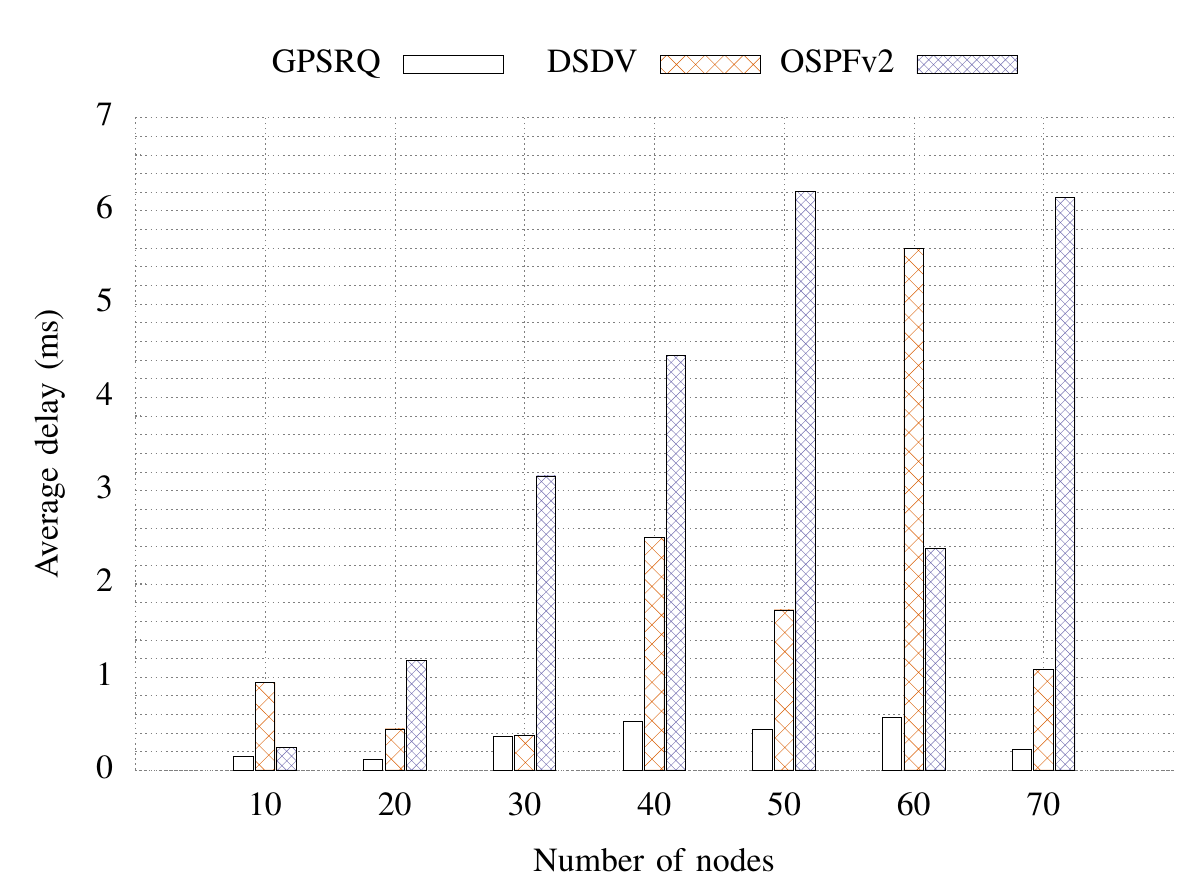}
  \end{center}
  \caption{\small Average delay}
  \label{fig:planar_delay} 
\end{figure}

\subsubsection{Path Length}
\label{sec:pathLength}

Fig.~\ref{fig:planar_pdr} shows the average number of hops. Data shows that GPSRQ recognizes paths to the destination which are not visible to DSDV and OSPFv2. More specifically, GPSRQ with the detection mechanism of returning the loop increases the number of forwarding instances, but due to scalable caching, GPSRQ is able to find the shortest feasible path to the destination.

\subsubsection{Time Delay}

Our results show that the scalable caching in GPSRQ has a significant impact on the average delay. As shown in Fig.~\ref{fig:planar_delay}, values are significantly lower when compared to DSDV and OSPFv2 due to the active measurements of link states and detection of returning loops to exclude those links that do not lead to the destination. The average delay for GPSRQ does not notably increase with the number of nodes, leading us to the conclusion that GPSRQ supports network scalability and robustness.

\subsubsection{The impact of GPSRQ $\beta$ parameter}
\label{section:effectOfBetaParameter}

 \begin{figure}[htbp]
  \begin{center}
  \includegraphics[width=0.4\textwidth]{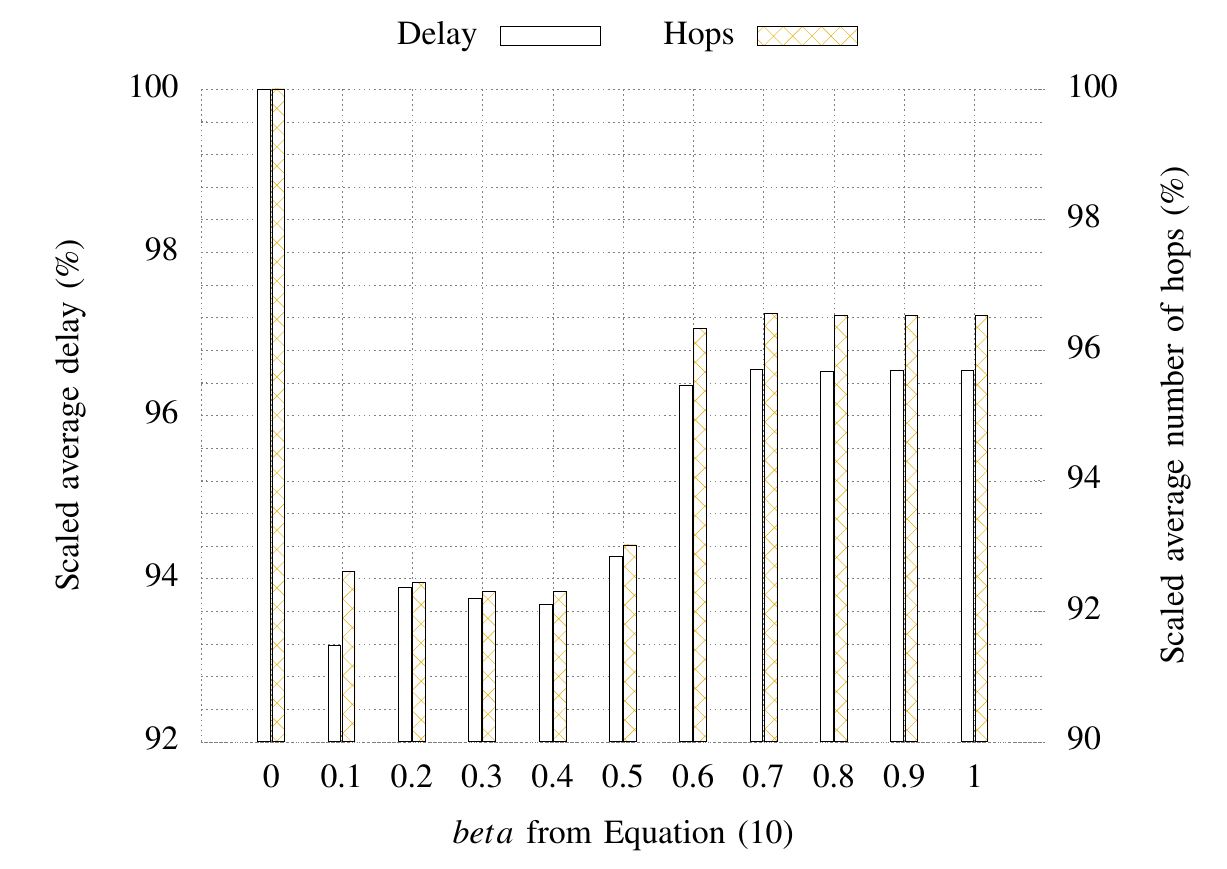}
  \end{center}
  \caption{\small Scaled average number of hops and scaled average delay for various values of $\beta$ (beta) from Equation~\eqref{eq:FQKD_QoS_Equation}.}
  \label{fig:planar_beta} 
\end{figure}

GPSRQ uses Equation~\eqref{eq:FQKD_QoS_Equation} to balance between the geographically shortest path and the path with the best performance. To show the impact of parameter $\beta$, we conducted simulations on static networks with 30, 40 and 50 randomly placed nodes, changing the values of $\beta$. Fig.~\ref{fig:planar_beta} shows that parameter $\beta$ which is used in greedy forwarding has a significant impact on the number of hops. Considering that GPSRQ implements recovery mode as a side algorithm which constantly seeks to forward packets to the destination using the right-hand rule, there are no significant differences in the PDR value for different values of $\beta$. However, $\beta$  directly affects the number of hops to the destination in greedy forwarding, which results in an increased delay and overall consumption of key material. Although a neighboring node may be geographically closer to the destination, routing without taking into consideration link performance ($\beta=1$) results in a returning loop and an increased delay. On the other hand, routing without taking into account the geographical distance forwards the packet further from the destination ($\beta=0$). However, with support from recovery mode, the packet reaches its destination in an increased number of hops and with a significantly greater delay. Intuitively, different classes of traffic should be served with different values of $\beta$ depending on the priority of delivery, which makes GPSRQ a flexible tool for traffic management and effective utilization of network resources.

\subsubsection{Impact of Caching Mechanism}
\label{section:effectOfPublicChannelMemory}

 \begin{figure}[htbp]
  \begin{center}
  \includegraphics[width=0.4\textwidth]{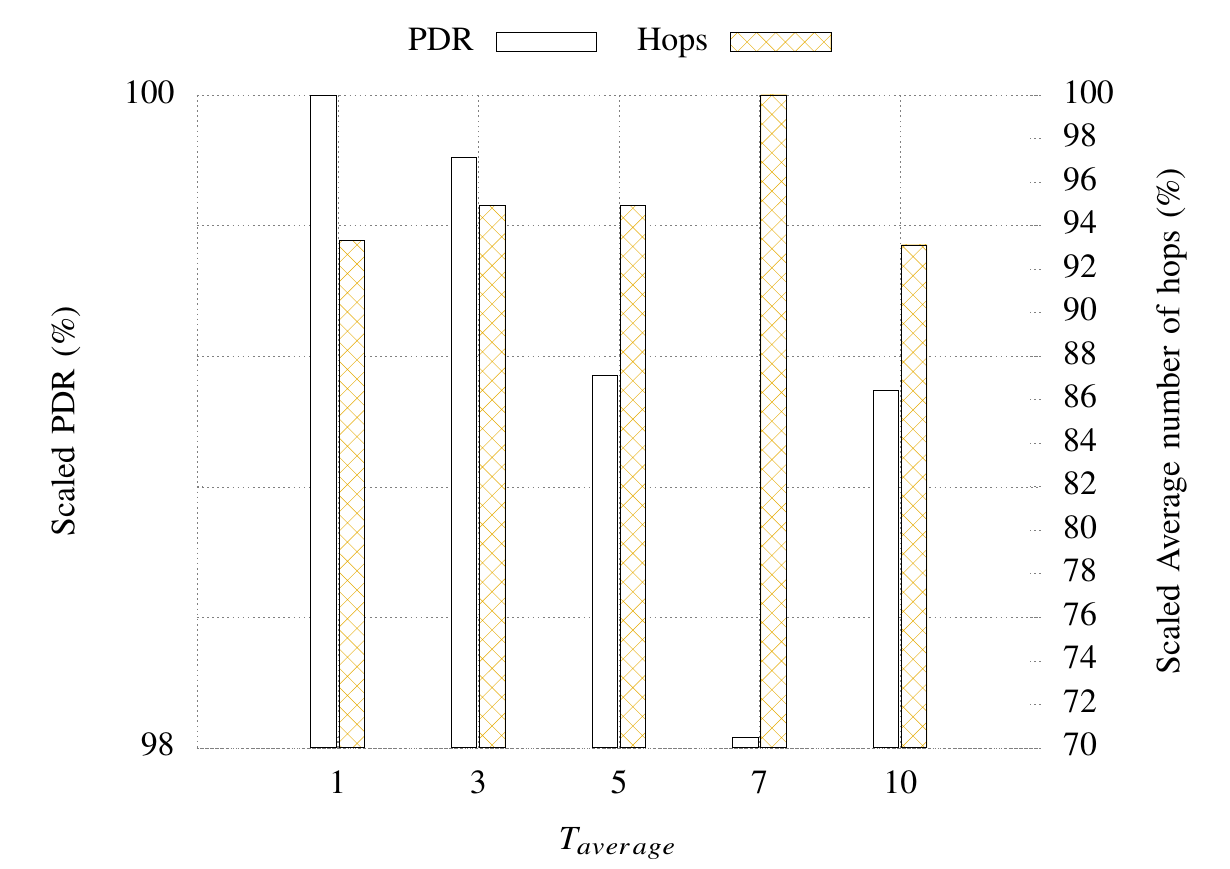}
  \end{center}
  \caption{\small Scaled average number of hops and scaled PDR for various values of $T_{average}$.}
  \label{fig:planar_cache_period}
\end{figure}

One of the key parameters of GPSRQ is the number of samples taken for calculating the average duration of key material establishment process in the long run, denoted as $T_{average}$ in Equation~\eqref{eq:publicChannel1}. $T_{average}$ defines the response time to variations of the public channel state and it is used to determine the validity of entries in the GPSRQ cache memory $T_{cache}$ as defined by Equation~\eqref{eq:t_cache}. $T_{average}$  affects the rate of emptying the internal node's cache which has a significant impact on the average delay. To demonstrate the impact of this parameter, we simulated random networks with 30, 40 and 50 nodes, changing the values of $T_{average}$. Fig.~\ref{fig:planar_cache_period} shows that the average number of hops grows with the $T_{average}$ since GPSRQ seeks alternative longer routes while PDR reduces as the validity of entries in internal node caches remains longer. In addition, we note that setting $T_{average}$ to the value of the time interval it takes to establish new key material (7 seconds in our simulations) results in a significant change of obtained values. The addition of new key material, defined by a small key charging rate, slightly changes the performance of the link. Therefore, frequent changes of cache memory records lead to an increased number of returning loops due to sudden changes in the availability of links.

 \begin{figure}[htbp]
  \begin{center}
  \includegraphics[width=0.4\textwidth]{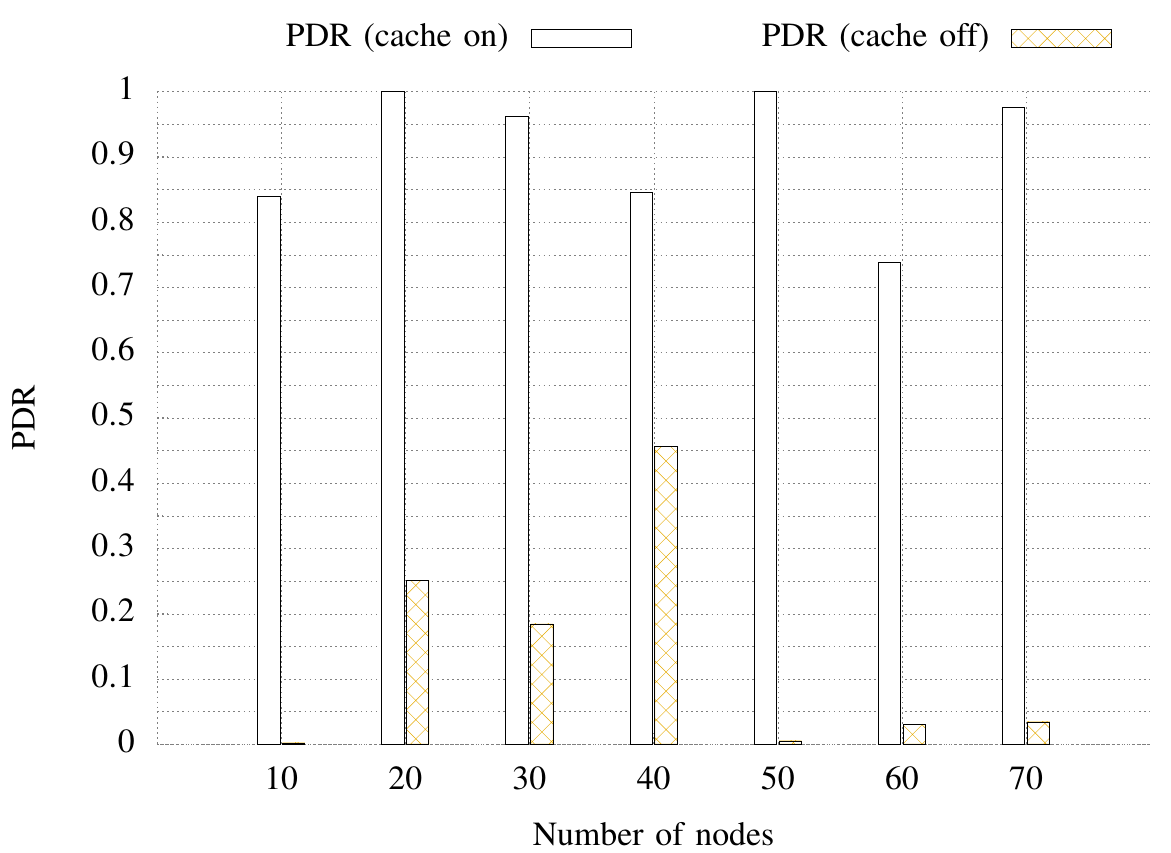}
  \end{center}
  \caption{\small The influence of the caching mechanism on PDR. $T_{average}=5$}
  \label{fig:planar_pdr_for_cache_on_cache_off} 
\end{figure}

We repeated the experiments with the cache mechanism excluded, where Fig.~\ref{fig:planar_pdr_for_cache_on_cache_off} shows the comparison of the PDR values. When the caching mechanism is turned off, the number of returning packets (packets with the value 2 of the loop field in the GPSRQ header) increases considerably. The key material is consumed on returning packets instead of data traffic which should be sent on the shortest path. Due to the larger traffic volume, the overall consumption of key material is substantially increased, which reduces the value of the PDR. Additionally, a noticeable increase in delay was noted due to longer routes.

\section{Conclusion}
\label{section:conclusion}

This paper provides a novel QoS model and routing protocol for QKD networks. The FQKD QoS model involves traffic classification at the ingress node based on prioritizing traffic into appropriate queues. It also implements additional waiting queues at a higher network layer to adapt to the dynamic nature of QKD networks. It defines specific metrics for public and quantum channels, an overall QKD link metric and $M_{thr}$ values to learn about the state of links which are more than one hop away. The GPSRQ routing protocol uses distributed geography and reactive routing to achieve high-level scalability. It is equipped with a caching mechanism and detection of returning loops, enabling forwarding while minimizing key material consumption. However, GPSRQ applications are limited to planar topologies only since geographic routing in networks with non-planar topologies are not able to quickly determine the shortest path, leading to unnecessary forwarding and increased consumption of scarce key material. Since QKD networks are limited to metropolitan scales~\cite{Ciurana2014,Wang2014,Peev2009}, and previously deployed QKD networks were deployed on planar topologies~\cite{Peev2009,Elliott2007,Xu2009,SasakiM20111,Wang2014}, we do not consider this restriction as a critical deficiency. Our simulation results show that GPSRQ outperforms most popular solutions in previously deployed testbeds. This is reflected particularly by minimizing the average delay and increasing PDR, which are the key parameters for operative use of the real-time application~\cite{Voznak2012a}.

The main contribution of this paper is providing novel metrics for determining the states of quantum and public channels as well as the overall state of QKD links, and providing a novel QoS model and routing protocol for QKD networks.

\section*{Acknowledgment}

\noindent
The research was funded by the SGS grant No. SP2018/59 "Networks and Communication Technologies for Smart Cities", VSB - Technical University of Ostrava, Czech Republic.
\appendices

\section{Modified QKD header and QKD command header}

\begin{table}[H]
\centering
\caption{Modified QKD header - description of fields}
\label{table:extendedQkdHeaderFields}
\resizebox{0.5\textwidth}{!}{%
\begin{tabular}{@{}lll@{}}
\toprule
Field & Length & Short Description \\ \midrule
Length & 32 bits & Total packet length in bytes \\
Message ID & 32 bits & Message ID \\ 
e & 4 bits & Type of encryption cipher used \\ 
a & 4 bits & Type of authentication algorithm used\\ 
z & 2 bits & Type of compression algorithm used\\ 
v & 2 bits & Version  \\
r & 2 bits & GPSRQ inRec indicator \\ 
l & 2 bits & GPSRQ loop indicator \\ 
Channel & 16 bits & QKD public channel ID \\
MaxDelay & 16 bits & Maximum tolerated time delay  \\
Timestamp & 16 bits &  \begin{tabular}[c]{@{}l@{}}Timestamp of the packet generation \\ at the ingress node \end{tabular} \\
Encryption Key ID & 32 bits & ID of key used for Encryption \\
Authentication Key ID & 32 bits & ID of key used for Authentication \\ 
Authentication-tag & 32 bits & Authentication tag \\ 
Payload & - & Data payload \\
\end{tabular}%
}
\end{table}

 \begin{figure}[H]
  \begin{center}
  \includegraphics[width=0.5\textwidth]{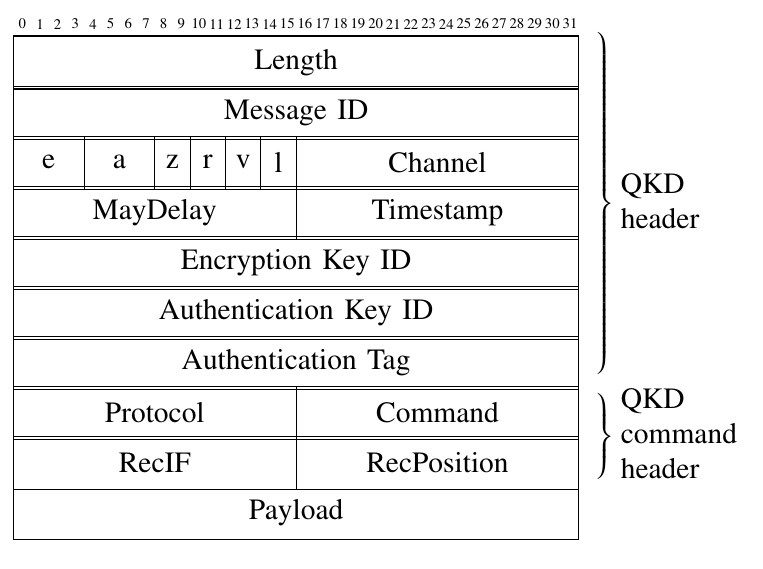}
  \end{center}  
\caption{\small Modified QKD header and QKD command header\\} 
\label{fig:qkdCommandHeader}  
\end{figure}

Table~\ref{table:extendedQkdCommandHeaderFields} and Table~\ref{table:extendedQkdHeaderFields} provide a short explanation of all fields in modified QKD headers and QKD command headers~\cite{Kollmitzer2010}.
 
\begin{table}[H]
\centering
\caption{QKD command header - Description of Fields}
\label{table:extendedQkdCommandHeaderFields}
\resizebox{0.5\textwidth}{!}{
\begin{tabular}{@{}lll@{}}
\toprule
Field & Length & Short Description \\ \midrule 
Protocol & 16 bits & Type of next header in the packet headers chain  \\
Command & 16 bits &  \begin{tabular}[c]{@{}l@{}}Key management sub-protocol operation command\\(LOAD NEW KEY and other.) \end{tabular} \\
RecIF & 16 bit & GPSRQ recIf field \\
RecPosition & 16 bit &  GPSRQ recPosition field\\
\end{tabular}%
}
\end{table}

\ifCLASSOPTIONcaptionsoff
  \newpage
\fi


%



%
%

\begin{small}
\bibliographystyle{IEEEtranTCOM}
\bibliography{library} 
\end{small}

\end{document}